\documentclass[amsmath,amssymb,aps,pra,reprint,superscriptaddress,footnoteinbib,twocolumn,longbibliography]{revtex4-2}
\usepackage{amsmath,amssymb,lmodern}
\usepackage[english]{babel}
\usepackage{bbold}
\usepackage{upgreek}
\usepackage{graphicx}
\usepackage{comment}
\usepackage{braket}
\usepackage{colortbl}
\usepackage{xr-hyper}
\usepackage[colorlinks]{hyperref}
\hypersetup{colorlinks,linkcolor=blue,citecolor=blue,urlcolor=black,final}
\usepackage[utf8]{inputenc}
\usepackage{xcolor}
\allowdisplaybreaks
\definecolor{cayenne}{rgb}{0.6, 0, 0}
\usepackage{dsfont}
\usepackage{setspace}
\usepackage{dcolumn}
\usepackage{newfloat}
\usepackage{cancel}
\usepackage{mathtools}
\usepackage{float}
\usepackage{bm}
\usepackage{datetime}
\usepackage{latexsym}
\usepackage{enumerate}
\usepackage{marvosym}
\usepackage{color,hyperref}
\usepackage{siunitx}

\begin{document}

\title{Nonlinear dynamics and mechanical frequency combs with a Meissner-levitated micromagnet}

\author{Y. Wang}
\thanks{These authors contributed equally to this work}
\affiliation{Department of Physics, Harvard University, Cambridge, Massachusetts 02138, USA}

\author{T. Madhavan}
\thanks{These authors contributed equally to this work}
\affiliation{Harvard John A. Paulson School of Engineering and Applied Sciences, Harvard University, Cambridge, Massachusetts 02138, USA}

\author{V. Wachter}
\thanks{These authors contributed equally to this work}
\affiliation{Institute for Complex Quantum Systems and Center for Integrated Quantum Science and Technology, Ulm University, Albert-Einstein-Allee 11, 89069 Ulm, Germany}

\author{J. D. Schaefer}
\thanks{Current address: Rigetti Computing, Berkeley, California, 94710, USA.}
\affiliation{Department of Physics, Harvard University, Cambridge, Massachusetts 02138, USA}

\author{A. NewRingeisen}
\affiliation{Quantum Science and Engineering, Harvard University, Cambridge, Massachusetts 02138, USA}

\author{Z. Wei}
\affiliation{Institute for Complex Quantum Systems and Center for Integrated Quantum Science and Technology, Ulm University, Albert-Einstein-Allee 11, 89069 Ulm, Germany}

\author{F. Fung}
\thanks{Current address: Applied Materials, Inc., Santa Clara, California, 95051, USA.}
\affiliation{Department of Physics, Harvard University, Cambridge, Massachusetts 02138, USA}

\author{B. A. Stickler}
\affiliation{Institute for Complex Quantum Systems and Center for Integrated Quantum Science and Technology, Ulm University, Albert-Einstein-Allee 11, 89069 Ulm, Germany}

\author{M. D. Lukin}
\email{lukin@physics.harvard.edu}
\affiliation{Department of Physics, Harvard University, Cambridge, Massachusetts 02138, USA}

%%%%%%%%%%%%%%%%%%%%%%%%%%%%%%%%%%%%%%%%%%%%%%%%%%
\begin{abstract} 
%%%%%%%%%%%%%%%%%%%%%%%%%%%%%%%%%%%%%%%%%%%%%%%%%%
Nonlinearities in multimode mechanical systems can give rise to rich dynamical phenomena with great potential for sensing applications and for future quantum experiments. We demonstrate that the coupled center-of-mass and rotational motion of a Meissner-levitated micromagnet offer a promising platform for nonlinear dynamics, combining low dissipation, magnetic tunability, strong intrinsic Duffing nonlinearities, and nonlinear intermodal couplings. We use this tunability to demonstrate the generation of a mechanical frequency comb in the micromagnet dynamics, realized by parametric excitation of two translational modes followed by cascaded nonlinear frequency mixing, which produces a phononic comb with tunable spacing. At large amplitudes, nonlinear coupling to a low-frequency librational mode of the magnet leads to parametric excitation and phase locking of that mode, generating a dense spectral fine structure at subharmonics of the drive. These results establish levitated micromagnets as a platform for nonlinear multimode mechanics, with potential applications in precision sensing and quantum-limited metrology.
\end{abstract}

\maketitle

\setlength{\abovecaptionskip}{4pt}
% %%%%%%%%%%%%%%%%%%%%%%%%%%%%%%%%%%%%%%%%%%%%%%%%%%
% \section{Introduction}
% %%%%%%%%%%%%%%%%%%%%%%%%%%%%%%%%%%%%%%%%%%%%%%%%%%
\begin{figure}
	\centering
	\includegraphics[width=\columnwidth]{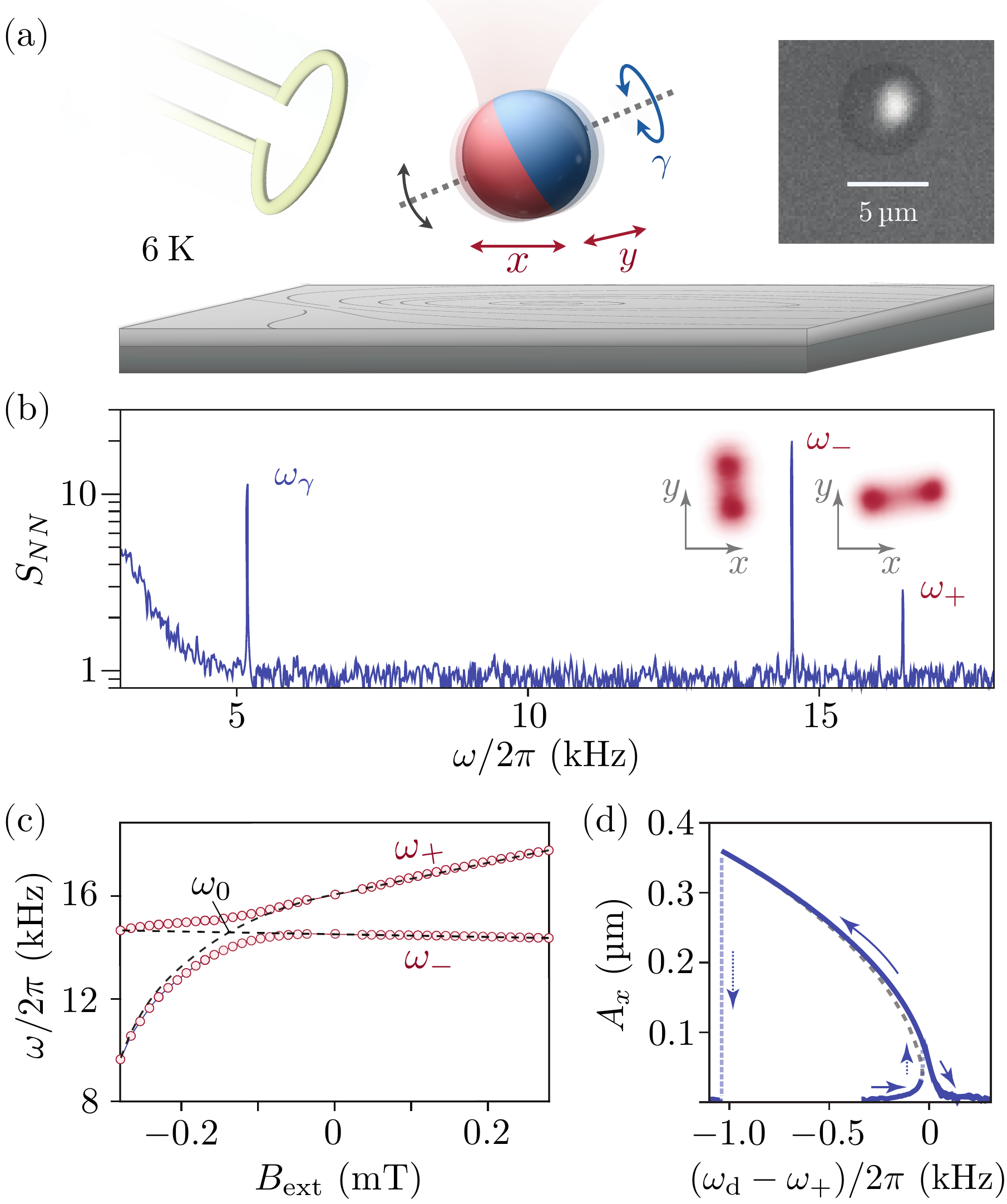}
	\caption{
	\textrm{(a)}~Schematic of the experimental setup. A micromagnet is levitated above a YBCO film in a cryostat at \SI{6}{\kelvin}. An \textit{in situ} coil drives the micromagnet, whose motion is detected using a focused laser beam. Inset: Optical image of the levitated micromagnet under broadband LED illumination. 
	\textrm{(b)}~Undriven photon-count-rate spectrum. From low to high frequencies, the peaks correspond to the $\gamma$ mode and the predominantly $y$-like and $x$-like hybrid modes. Insets: Spatial scans of the driven $x$- and $y$-like modes.
	\textrm{(c)}~Measured hybridized eigenfrequencies $\omega_+$ and $\omega_-$ as functions of $B_\mathrm{ext}$. The dashed black curves show the inferred uncoupled frequencies $\omega_x$ and $\omega_y$, which cross at the common frequency $\omega_0$ (see SI). 
	\textrm{(d)}~Nonlinear response of the measured $x$-projected amplitude $A_x$ of the predominantly $x$-like mode at a drive power of $-20$~dBm as a function of the detuning $\omega_\mathrm{d}-\omega_+$. Arrows indicate the sweep direction, dashed blue lines mark discontinuous amplitude jumps, and the dashed gray curve denotes the unstable solution obtained from Eq.~\eqref{eq:duffing_amp} in SI~\cite{seeSI}.}
	\label{fig1:schematic} 
\end{figure}
 
Nonlinear systems constitute a key resource in both classical and quantum processes and applications~\cite{strogatz2024nonlinear,haakequantum,dykman2012fluctuating}. Classically, nonlinearities support self-sustained time-periodic dynamics, bifurcations and criticality, and synchronization phenomena with various applications~\cite{guckenheimer1983nonlinear, li2022all, wu2026nonlinear, wang2026exceptional,deng2023amplifying,zhang2026cusp,margiani2026activated}. In the quantum regime, nonlinearities are required for the creation and detection of non-classical states of otherwise harmonic systems~\cite{fein2019quantum,bild2023schrodinger,panda2024measuring,pedalino2026probing, hofheinz_2008_nature, vion_2002_science}. In the mechanical domain, 
the evolution of nanoscale-to-microscale mechanical objects in nonlinear potentials offers a route toward the next generation of macroscopic quantum interference experiments~\cite{scala2013matter,bateman2014near,stickler2018probing,roda-llordes2024macroscopic}.
 
One striking manifestation of nonlinear classical dynamics is the emergence of steady-state trajectories that exhibit phase-coherent spectra composed of many regularly spaced sidebands. Well known in the optical domain~\cite{bartels_2009_science, delhaye_2007_nature}, this phenomenon also occurs in mechanical systems, resulting in the so-called mechanical frequency combs~\cite{ochs_2022_PRX,fu2025sideband,deJong_2023_NatComms,ganesan_2017_PRL,czaplewski_2018_PRL,ye2025magnetostrictive,wu_2025_PRL}. These mechanical combs can arise in multimode systems through nonlinear mode coupling and, in analogy to optical frequency combs~\cite{diddams_2004_Science,picque2019frequency,jones2000carrier}, they could leverage frequency-stable mechanical spectra for compact sensing and coherent transduction applications~\cite{ganesan_2019_SciRep, ganesan_2019_IEEE, li_2024_Frontiers,cao_2014_PRL}. 
Yet stable mechanical comb generation remains challenging, because the same nonlinearities that produce phase-coherent spectra can also destabilize the underlying motion through amplitude-dependent frequency shifts, nonlinear dispersion, and unwanted intermodal coupling. Robust comb formation typically requires carefully engineered mode structures, strong driving, and low dissipation~\cite{ganesan_2017_PRL,czaplewski_2018_PRL,han_2022_PRL, goryachev_2020_PRR}. 

Magnetically levitated particles provide a promising platform for exploring nonlinear multimode dynamics~\cite{gutierrez2023superconducting,latorre2022chip,janse2024characterization,perdriat2023planar,cirio2012quantum,fuwa2023ferromagnetic,hofer2023high}. Micromagnets can be stably suspended above superconducting surfaces~\cite{gieseler_2020_PRL,ahrens_2025_PRL,ahrens2025observation,timberlake2019acceleration,vinante2022levitated,jose2025cryogenic,ren2026field,prat2017ultrasensitive}, creating a setup that combines exceptionally low dissipation with tunable magnetic trapping landscapes and intrinsically nonlinear rotational dynamics, including spin-rotation coupling~\cite{rusconi2017linear,rusconi2022spin, wachter2021optical,wachter2026gyroscopically,perdriat2024rotational,sato2023gyromagnetic,jin2024quantum,ma2021torque,ahrens2025observation,delord2020spin}. Their strongly coupled translational and librational modes remain continuously observable via optical readout and controllable via magnetic fields, enabling access to nonlinear dynamics that are difficult to assess in other mechanical systems. While these traits make levitated micromagnets an attractive setting for studying nonlinear phenomena, including time-crystalline dynamics~\cite{wilczek2012quantum,choi2017observation, sacha2018time,zaletel2023colloquium,moon2026sensing} and mechanical frequency combs, nonlinear multimode dynamics have remained largely unexplored in magnetically trapped objects. 

In this Letter, we show that Meissner-levitated micromagnets provide a magnetically tunable platform for engineering strongly nonlinear multimode dynamics, resulting in a distinct dynamical regime of mechanical frequency combs. Continuous optical monitoring during a drive-frequency sweep resolves the comb-formation process, ultimately producing a full comb with more than 60 evenly spaced spectral teeth. The comb formation can be understood in terms of intrinsic nonlinear multimode interactions and tunable hybridization of two translational modes, which generate a primary comb, together with coupling to a librational mode that gives rise to the full comb and its additional spectral structure. This mechanism suggests a universal route toward phononic frequency-comb generation. The resulting mechanical comb exhibits tunable spacing together with stable nonlinear multimode dynamics, opening new opportunities for exploiting strong nonlinearities in sensing~\cite{deng2023amplifying} and magnetometry~\cite{lassagne_2011_PRL,ji2025levitated,ni2025microscopic} with magnetically levitated systems.
 
Our experimental platform consists of a spherical NdFeB micromagnet of radius \SI{1.8}{\micro\meter}, levitated \SI{4.4}{\micro\meter} above the surface of a type-II superconductor in high vacuum ($\leq10^{-6}\,\mathrm{mbar}$) at \SI{6}{\kelvin}, as shown in Fig.~\ref{fig1:schematic}(a) and further specified in Supplementary Information~\cite{seeSI}.
Levitation results from the combined action of repulsive Meissner screening currents and the attractive interaction with magnetic flux trapped and pinned as the superconductor is field-cooled through its transition at \SI{87}{\kelvin}~\cite{kordyuk1998magnetic,wei2026meissner,wang2026exploring}. The motion of the micromagnet is read out optically via intensity modulation of reflected light from a \SI{637}{\nano\meter} probe laser with a power of \SI{0.5}{\nano\watt}. The resulting photon-count spectrum $S_{NN}$ exhibits six distinct mechanical resonances, three of which are shown in Fig.~\ref{fig1:schematic}(b), indicating stable confinement of all three translational and three rotational degrees of freedom. The observed mode structure is captured by the effective trapping potential derived in SI~\cite{seeSI}. In particular, the lowest-frequency resonance, at $\omega_{\gamma}/2\pi=\SI{5.16}{\kilo\hertz}$, corresponds to libration about the magnetic-dipole axis ($\gamma$ mode). The other two resonances occur at $\omega_{+}/2\pi=\SI{16.44}{\kilo\hertz}$ and $\omega_{-}/2\pi=\SI{14.53}{\kilo\hertz}$, with corresponding quality factors $Q_+=2.95(8)\times10^5$ and $Q_-=2.0(3)\times10^5$. Spatial scans of the optical response [insets of Fig.~\ref{fig1:schematic}(b)] associate the upper and lower resonances predominantly with motion along two in-plane directions, which we denote by $x$ and $y$, respectively. Both modes also contain strong librational components, namely out-of-plane tilting of the magnetic dipole moment for the $x$-like mode and in-plane rotation for the $y$-like mode; however, for brevity, we refer to them below by their predominant translational character. 

Applying an external homogeneous static magnetic field $B_\mathrm{ext}$ parallel to the superconducting surface and approximately along the $x$ direction tunes both resonance frequencies over a wide range. As $B_\mathrm{ext}$ is varied, the mode frequencies $\omega_\pm$ exhibit an avoided crossing [Fig.~\ref{fig1:schematic}(c)], identifying the corresponding modes as hybridized eigenmodes of two linearly coupled in-plane degrees of freedom. We denote the corresponding effective uncoupled frequencies of the underlying $x$ and $y$ modes by $\omega_x$ and $\omega_y$, which coincide at $\omega_0$ at the center of the avoided crossing. A fit yields a linear intermodal coupling rate of $g/2\pi=\SI{511(10)}{\hertz}$. At the bias field $B_\mathrm{ext}=\SI{37}{\micro\tesla}$ used throughout the remainder of this work, the upper and lower hybridized eigenmodes are predominantly $x$-like and $y$-like, respectively. 

\begin{figure}[t!]
	\centering
	\includegraphics[width=\columnwidth]{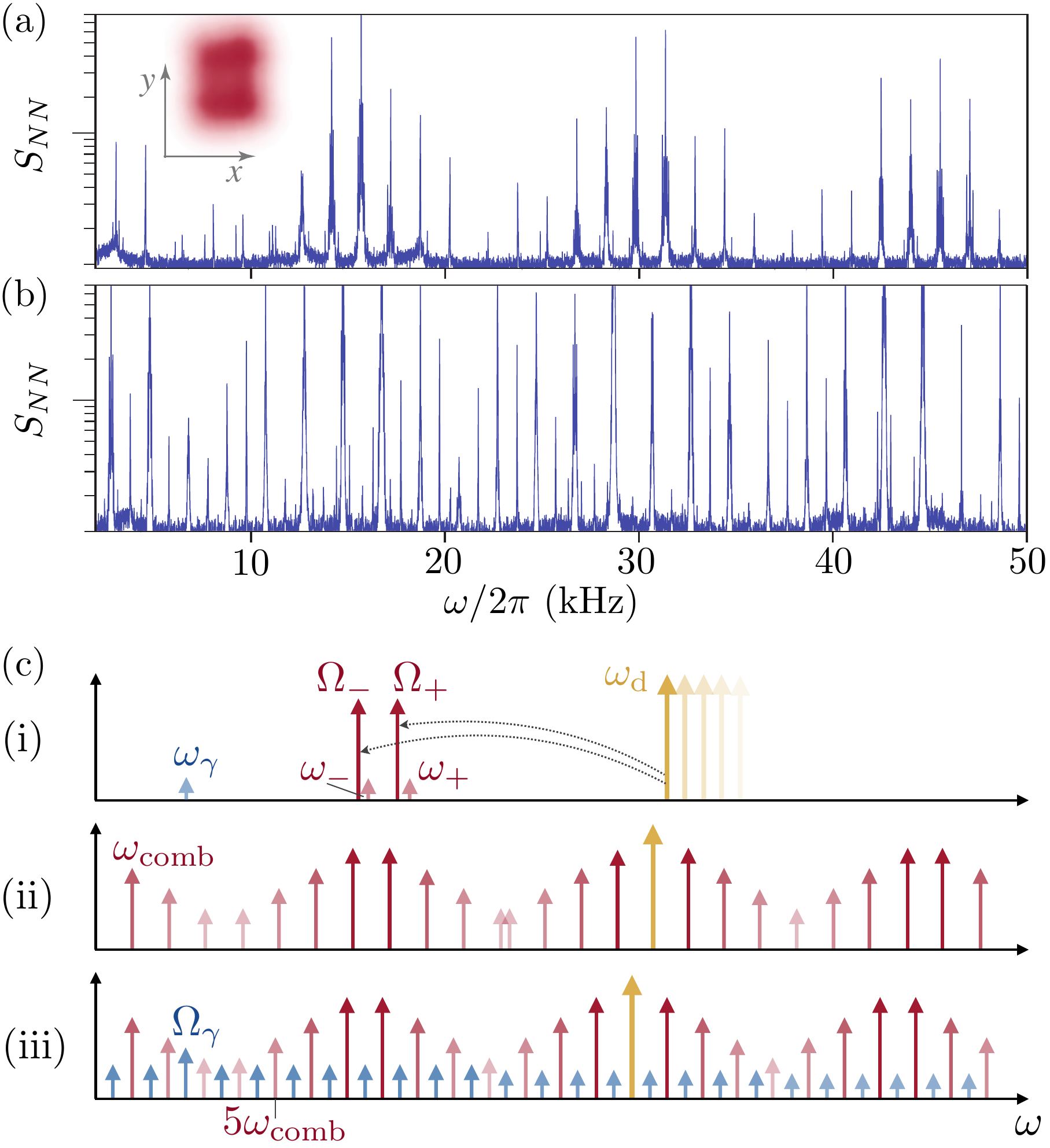}
	\caption{
	\textrm{(a)}~Photon-count-rate spectrum under an $8~\mathrm{dBm}$ drive. Inset: spatial scan of driven motion. 
	\textrm{(b)}~Photon-count-rate spectrum for the same drive power at $\omega_{\mathrm{d}}/2\pi=\SI{30.06}{\kilo\hertz}$ during the downward sweep in Fig.~\ref{fig3:gamma}. 
	\textrm{(c)}~Schematic illustration of the mechanical frequency comb formation process: (i)~off-resonant driving evolving into parametric excitation of the carrier frequencies as $\omega_\mathrm{d}$ is decreased, (ii)~parametric driving and cascaded sideband generation resulting in the primary comb, (iii)~$\gamma$-mode excitation and full-comb formation.}
	\label{fig2:comb} 
\end{figure}

The magnetic-field dependence of the mechanical mode frequencies enables their parametric modulation. We use an \textit{in situ} coil to apply an oscillating magnetic field at angular frequency $\omega_\mathrm{d}$. The oscillating field also provides an additive coherent drive, manifested as a pronounced spectral peak at $\omega_{\mathrm{d}}$ [Fig.~\ref{fig2:comb}(a)]. When $\omega_\mathrm{d}$ is tuned near $\omega_+$, this additive component excites the predominantly $x$-like mode to an $x$-projected amplitude $A_x$ sufficiently large to reveal the intrinsic anharmonicity of the trapping potential. Figure~\ref{fig1:schematic}(d) shows the resulting softening Duffing response, characterized by a bent resonance curve and hysteresis between upward and downward frequency sweeps. From the measured amplitude dependence of the resonance frequency, we extract an effective Duffing coefficient $\lambda_x/(2\pi)^2=-1.5(3)\times 10^{22}~\mathrm{Hz^2/m^2}$~\cite{seeSI}. An analogous measurement of the predominantly $y$-like lower eigenmode yields the corresponding Duffing coefficient $\lambda_y/(2\pi)^2=-0.5(1)\times 10^{22}~\mathrm{Hz^2/m^2}$. 

%\textit{Observed frequency-comb dynamics.---} 
Strong driving reveals nonlinear multimode dynamics in the levitated micromagnet, including the emergence of a mechanical frequency comb. To access this regime, we apply a drive of $8~\mathrm{dBm}$ and adiabatically sweep its angular frequency downward from well above $\omega_++\omega_-$, where the measured response is weak. As $\omega_\mathrm{d}$ approaches the sum of the independently measured linear-response eigenfrequencies, $\omega_\mathrm{d}\approx \omega_++\omega_-$, the spectral response near both modes increases sharply and regularly spaced peaks emerge, as seen in Fig.~\ref{fig2:comb}(a). We refer to this initial comb state, associated predominantly with the $x$-like and $y$-like modes, as the \textit{primary comb}, distinguishing it from the broader full-comb state described below. The corresponding time-averaged spatial distribution of the micromagnet exhibits a nearly rectangular envelope [inset of Fig.~\ref{fig2:comb}(a)].

\begin{figure*}[t!]
	\centering
	\includegraphics[width=\textwidth]{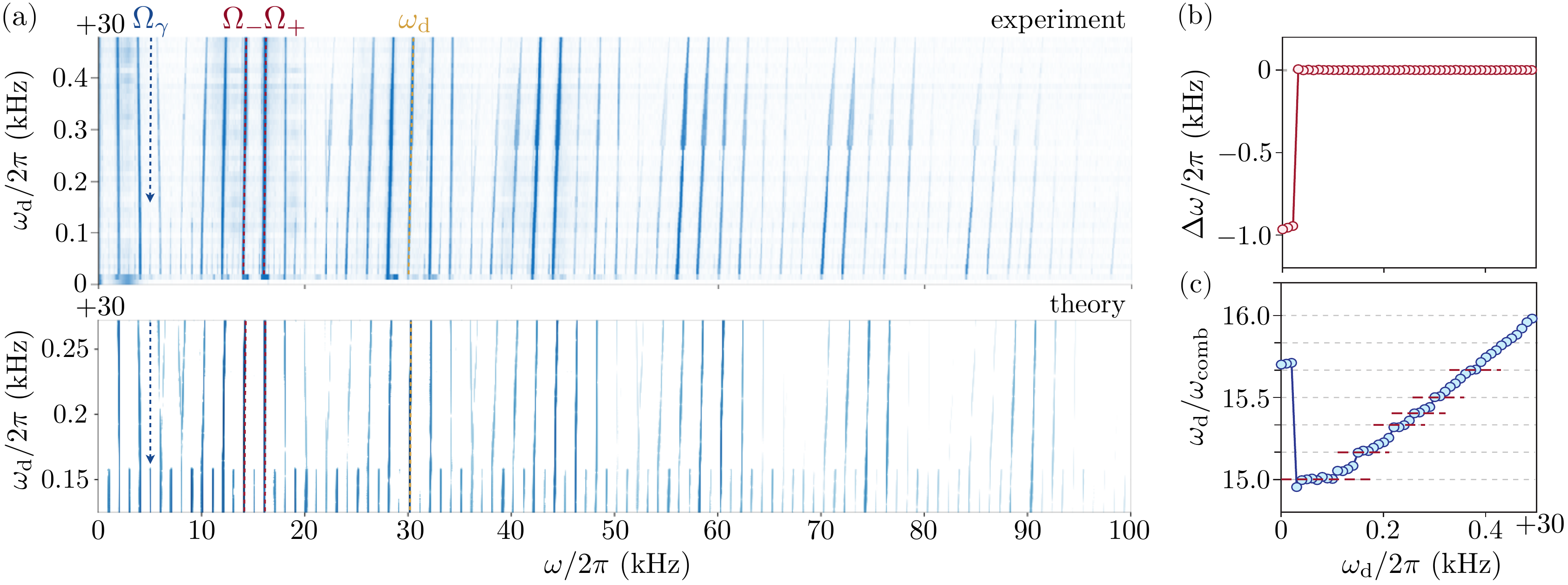}
	\caption{
	\textrm{(a)}~Upper panel: measured photon-count spectra during the downward sweep of the drive frequency $\omega_\mathrm{d}$. Dashed lines indicate the $\gamma$ mode with effective frequency $\Omega_\gamma$ (blue), the predominantly $y$-like (red) and $x$-like (red) carrier tones, and the drive (yellow). Lower panel: simulated sweep spectrum obtained from the extended nonlinear model in Eqs.~\eqref{eq:eom_xy} and \eqref{eq:eom_gamma}, showing qualitative agreement with the experimental data. Phenomenological simulation parameters are detailed in SI~\cite{seeSI}. 
	\textrm{(b)}~Measured frequency mismatch $\Delta\omega=\omega_{\mathrm{d}}-(\Omega_{+}+\Omega_-)$ versus $\omega_{\mathrm{d}}$ throughout the frequency sweep. 
	\textrm{(c)}~Corresponding ratio $\omega_\mathrm{d}/\omega_\mathrm{comb}$ of the drive frequency to the primary-comb spacing versus $\omega_{\mathrm{d}}$ during the same sweep. Dashed red lines mark the ratios $15+1/6$, $15+1/3$, $15+2/5$, $15+1/2$, $15+2/3$ from bottom to top.}
	\label{fig3:gamma} 
\end{figure*}

The upper panel of Fig.~\ref{fig3:gamma}(a) resolves the evolution of the primary comb during the downward sweep. After its onset, two dominant spectral peaks remain close to the linear-response eigenfrequencies $\omega_\pm$. We denote their drive-dependent frequencies by $\Omega_\pm$. Over a finite interval of the sweep, their sum tracks the drive frequency within experimental resolution, $\Omega_++\Omega_-=\omega_\mathrm{d}$, whereas outside this interval, $\Omega_\pm$ return toward $\omega_\pm$ [Fig.~\ref{fig3:gamma}(b)]. The additional spectral peaks form regularly spaced sequences with spacing $\omega_\mathrm{comb}\equiv|\Omega_+-\Omega_-|$, which varies continuously during the sweep.

Throughout the primary-comb regime, the ratio $\omega_\mathrm{d}/\omega_\mathrm{comb}$ increases approximately linearly with $\omega_\mathrm{d}$, while weak slope changes occur near several rational frequency ratios marked in Fig.~\ref{fig3:gamma}(c). Near $\omega_\mathrm{d}/2\pi \approx 30.11\,\mathrm{kHz}$, the ratio reaches the integer value 15 and develops a plateau spanning a drive-frequency interval of approximately $\Delta\omega_\mathrm{d}/2\pi\approx \SI{70}{\hertz}$. At the same point, several peaks in the observed spectrum merge.

Upon further decreasing the drive frequency to $\omega_\mathrm{d}/2\pi \approx 30.09\,\mathrm{kHz}$, the spectrum enters a qualitatively distinct regime, exhibiting two characteristic frequency scales: the primary-comb spacing $\omega_\mathrm{comb}$ remains visible, while additional spectral peaks emerge between neighboring primary-comb teeth, producing a finer spacing of approximately $\omega_\mathrm{comb}/2\approx 2\pi\times\SI{1}{\kilo\hertz}$ [Fig.~\ref{fig2:comb}(b)]. The resulting spectrum contains more than 60 resolved, evenly spaced teeth above the noise floor. We thus call this regime the \textit{full comb}. At $\omega_{\mathrm{d}}/2\pi\approx\SI{30.06}{\kilo\hertz}$, additional weak spectral peaks emerge between the existing comb teeth, reducing the smallest resolved spacing to $2\pi\times\SI{0.5}{\kilo\hertz}$. For $\omega_\mathrm{d}/2\pi < \SI{30.03}{\kilo\hertz}$, the relation $\Omega_++\Omega_-=\omega_\mathrm{d}$ breaks down and the well-resolved comb state vanishes [Fig.~\ref{fig3:gamma}(a)]. 

The observed hierarchy of comb states is captured by the three stages (i)–(iii) illustrated in Fig.~\ref{fig2:comb}(c). The primary comb can be understood to arise from the dynamics of two linearly coupled Duffing oscillators, 
\begin{subequations}\label{eq:eom_xy}
	\begin{align}
		\ddot{x}=&-\Gamma_x\dot{x}-\omega_x^2x - 2 g\omega_0 y-\lambda_x x^3+f_{\rm dr}(t),\\
		\ddot{y}=&-\Gamma_y\dot{y}-\omega_y^2y- 2 g\omega_0 x -\lambda_y y^3,
	\end{align}
\end{subequations}
where only the $x$ coordinate is driven through the combined coherent and parametric drive
\begin{equation}
	f_{\rm dr}(t) = \left (\eta + \varepsilon x \right ) \cos(\omega_\mathrm{d}t).
\end{equation}
Here, $\Gamma_{x,y}$ denote the effective damping rates, while $\eta$ and $\varepsilon$ quantify the coherent and parametric magnetic-drive components, respectively. The approximate hybrid eigenfrequencies of the linear part of the coupled equations are $\omega_{\pm}\approx\omega_{x,y}\pm2g^2\omega_0^2/\omega_{x,y}(\omega_{x}^2-\omega_{y}^2)$.
Near the sum-frequency resonance, $\omega_\mathrm{d}\approx\omega_++\omega_-$, the parametric drive couples the two hybrid modes through nondegenerate three-wave mixing~\cite{eichler2023classical}, while the Duffing nonlinearity mediates four-wave mixing. The coherent drive is far detuned from either hybrid-mode resonance and therefore does not contribute resonantly to the slow-envelope dynamics of the eigenmodes. Projecting Eq.~\eqref{eq:eom_xy} onto the hybrid-mode basis and retaining the dominant resonant terms within a rotating-wave approximation shows that, at exact sum-frequency resonance, the zero-amplitude state becomes unstable when 
\begin{equation}
	\varepsilon^2>4\omega_+\omega_-\left[(\Gamma_x-\Gamma_y)^2+\Gamma_x\Gamma_y\left(\dfrac{\omega_+^2-\omega_-^2}{2g\omega_0}\right)^2\right].
	\label{eq:instability_threshold}
\end{equation}
Above this instability threshold, both hybrid modes acquire finite amplitudes [stage (i)], and their growth saturates when dissipation and nonlinear frequency shifts balance the parametric gain, as detailed in SI~\cite{seeSI}. 

For finite detuning $\Delta=\omega_\mathrm{d}-(\omega_++\omega_-)$, the existence and stability of locked finite-amplitude solutions depend on the dynamics of the carrier phases $\varphi_\pm(t)$ in frames rotating at $\omega_\pm$. The phase mismatch between the drive and the parametrically generated carrier pair, $\Phi(t) = \varphi_+(t)+\varphi_-(t)+\Delta\,t$, obeys Adler-type dynamics specified in SI~\cite{seeSI}. Within the corresponding locking range, these dynamics admit a stable fixed point, $\dot{\Phi}=0$, locking the sum of the carrier phases to the drive. Defining the drive-dependent carrier frequencies as $\Omega_\pm=\omega_\pm-\dot{\varphi}_\pm$,
the fixed-point condition enforces $\Omega_++\Omega_-=\omega_\mathrm{d}$, as observed in Fig.~\ref{fig3:gamma}(b). While the sum phase is locked by the parametric interaction, the relative phase remains unconstrained. For incommensurate $\Omega_+$ and $\Omega_-$, the resulting motion quasiperiodically explores a dense set of relative phases, consistent with the nearly rectangular trajectory in the inset of Fig.~\ref{fig2:comb}(a). 

Beyond the leading rotating-wave treatment, these phase-correlated carrier tones seed the nonlinear frequency conversion underlying the primary comb. As their amplitudes increase, the negative Duffing nonlinearities contribute amplitude-dependent downward shifts of $\Omega_\pm$ relative to $\omega_\pm$ and mediate four-wave mixing between the carriers. Repeated mixing processes generate progressively higher-order sidebands, producing a cascade of combination tones of the form $j\Omega_++k\Omega_-$, with $j,k\in\mathbb{Z}$. Within each family of fixed $j+k$, adjacent spectral components are separated by the primary-comb spacing $\omega_\mathrm{comb}=|\Omega_+-\Omega_-|$. The cascaded four-wave mixing generates the dense sideband ladders around the carrier tones and the drive frequency [stage (ii)]. During the downward sweep, the accompanying drive-dependent nonlinear frequency shifts continuously tune $\Omega_\pm$ and hence the primary-comb spacing $\omega_\mathrm{comb}$. Consequently, combination tones generated through distinct mixing pathways can become degenerate at selected rational values of $\omega_\mathrm{d}/\omega_\mathrm{comb}$~\cite{seeSI}. Their overlap can enhance or redistribute spectral weight across the comb, accounting for the spectral rearrangements observed in Figs.~\ref{fig3:gamma}(a) and (c). Note that the nonlinear system described in Eq.~\eqref{eq:eom_xy} can also be driven into a limit-cycle regime, as recently observed in Ref.~\cite{margiani2026activated}.

To model dynamical locking of $\omega_\mathrm{comb}$ and the formation of the full comb, we extend Eq.~\eqref{eq:eom_xy} by including $\gamma$ as a third nonlinear oscillator. The $\gamma$ mode couples to $x$ both linearly, with effective coupling coefficient $g_\gamma$, and nonlinearly through the leading cubic interactions proportional to $x\gamma^2$ and $x^2\gamma$ in the trapping potential~\cite{seeSI}. Its equation of motion is
\begin{equation}
    \ddot{\gamma} =- \Gamma_\gamma \dot{\gamma} - \omega_\gamma^2\gamma  - \lambda_\gamma \gamma^3  -g_\gamma x - 2 \chi x\gamma -  \zeta x^2,
    \label{eq:eom_gamma}
\end{equation}
where $\Gamma_\gamma$ and $\lambda_\gamma$ are the damping rate and Duffing coefficient, while $\chi$ and $\zeta$ denote effective nonlinear coupling coefficients. The term $2\chi x\gamma$ parametrically modulates the $\gamma$-mode frequency, whereas $\zeta x^2$ provides an additive nonlinear drive. The corresponding reciprocal $\gamma$-dependent forces acting on $x$ are included in the extended model, as specified in SI~\cite{seeSI}. A sufficiently strong primary-comb component of $x(t)$ can therefore parametrically excite the $\gamma$ mode when $n\omega_\mathrm{comb}\approx2\omega_\gamma$, $n\in\mathbb{N}$. This resonance condition is approximately satisfied for $n=5$ at $\omega_{\mathrm{d}}/2\pi\approx30.09~\text{kHz}$, coinciding with the onset of the full comb in Fig.~\ref{fig3:gamma}(a). Above threshold, the $\gamma$ mode develops a phase-locked oscillation at half the resonant modulation frequency $\Omega_\gamma=5\omega_\mathrm{comb}/2\approx\omega_{\mathrm{d}}/6$. Nonlinear mixing of this oscillation with the primary comb generates the additional sidebands between the primary-comb peaks with a characteristic spacing of approximately $\omega_{\mathrm{d}}/30$ [stage (iii)], providing a natural explanation for the finer structure [Fig.~\ref{fig2:comb}(b)]. 

Numerical simulations of the extended version of Eqs.~\eqref{eq:eom_xy} and \eqref{eq:eom_gamma} during a downward sweep qualitatively reproduce the progression from the primary comb and sideband merging to $\gamma$-mode excitation and the fine structure of the full comb [lower panel in Fig.~\ref{fig3:gamma}(a)]. Once excited, the $\gamma$ mode renormalizes the translational carrier frequencies through its nonlinear coupling to $x$.
Together with sum-phase locking, this feedback can therefore modify the evolution of $\omega_\mathrm{comb}$ during the downward drive-frequency sweep and stabilize $\omega_\mathrm{comb}\approx\omega_\mathrm{d}/15$ over a finite interval, yielding a plateau similar to that observed in Fig.~\ref{fig3:gamma}(c). These results support nonlinear coupling to the $\gamma$ librational mode as a plausible common mechanism for the full-comb substructure and the plateau in $\omega_\mathrm{d}/\omega_\mathrm{comb}$. We attribute the weaker substructure near $\omega_{\mathrm{d}}/2\pi\approx\SI{30.06}{\kilo\hertz}$ to thermal-noise-induced sidebands of the predominantly $x$-like mode~\cite{seeSI}, whose nonlinear mixing produces secondary comb families with reduced spacing~\cite{huber_2020_PRX}. Upon leaving the phase-locking range, the stable Adler fixed point disappears. The mode pair consequently loses sum-phase locking, resulting in the observed termination of the comb and the relaxation of the carrier frequencies toward the linear eigenfrequencies $\omega_\pm$ [Fig.~\ref{fig3:gamma}(b)].

\begin{figure}[t!]
    \centering
    \includegraphics[width=\columnwidth]{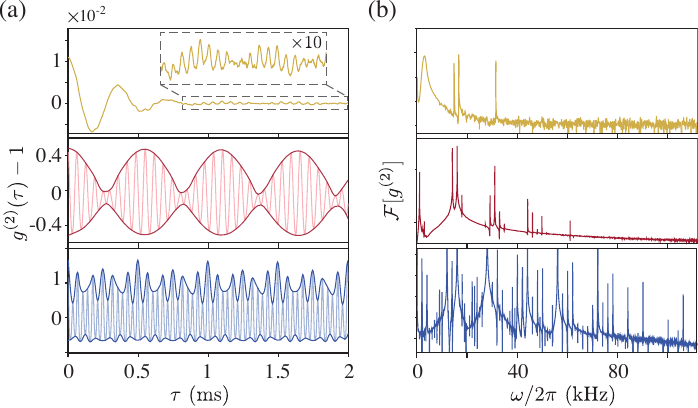}
    \caption{
    \textrm{(a)}~Second-order correlation functions $g^{(2)}(\tau)$ and \textrm{(b)}~their corresponding spectra. Panels from top to bottom are representative examples in the frequency sweep shown in Fig.~\ref{fig3:gamma}(a) in the stages (i) off-resonance driving (yellow), (ii) primary-comb regime (red), and (iii) full-comb regime (blue). Solid lines highlight envelopes of $g^{(2)}(\tau)$ in (a). The inset in the top panel of (a) is magnified $\times10$ vertically for visibility.
    }
    \label{fig4:coherence} 
\end{figure}

We probe the temporal coherence of the frequency comb through the second-order correlation function $g^{(2)}(\tau)$ of the optical readout. Because photon shot noise limits direct reconstruction of the motion, we evaluate $g^{(2)}(\tau)$ from photon coincidences separated by a delay $\tau$ and accumulated over \SI{100}{\second} at representative stages of the frequency sweep. Figure~\ref{fig4:coherence} shows the resulting (a) correlation functions and (b) their corresponding spectra. For far-off-resonant driving, $\omega_{\mathrm{d}}\gg\omega_++\omega_-$ [stage~\textrm{(i)}], $g^{(2)}(\tau)$ exhibits no persistent periodic structure, shown as yellow traces, while its spectrum resolves the hybrid-mode resonances $\omega_\pm$ and the off-resonant drive. Note that the rapidly decaying oscillation at short delays is a result of incoherent measurement noise, which appears as the broad peak near 2~kHz in the spectrum.
In the primary-comb regime [stage~\textrm{(ii)}], $g^{(2)}(\tau)$ develops a pronounced beating pattern with a period of approximately \SI{553}{\micro\second}, highlighted by the red envelope. This regular modulation indicates the emergence of persistent periodic correlations in the parametrically driven multimode state. In the full-comb regime [stage~\textrm{(iii)}], the structured correlations persist with a repetition period of approximately \SI{499}{\micro\second}, shown by the blue traces, consistent with a phase-locked nonlinear steady state. These time-domain correlations distinguish the comb regimes from off-resonant thermal motion and provide evidence for temporally stable nonlinear dynamics.

% %%%%%%%%%%%%%%%%%%%%%%%%%%%%%%%%%%%%%%%%%%%%%%%%%%
% \section{Discussion}
% %%%%%%%%%%%%%%%%%%%%%%%%%%%%%%%%%%%%%%%%%%%%%%%%%%
This Letter establishes Meissner-levitated micromagnets as a versatile platform for nonlinear multimode mechanics. Parametric excitation of the two hybridized in-plane translational modes generates a primary frequency comb with tunable spacing through cascaded nonlinear mixing, while nonlinear intermodal coupling to a librational mode produces the full comb with additional spectral structure. The resulting nonlinear motion exhibits periodic dynamics at subharmonics of the drive, reflected in the formation of sharp and regularly spaced spectral peaks. Such dense, tunable spectra could enable multifrequency sensing of forces, torques, and magnetic fields~\cite{deng2023amplifying, ni2025microscopic, ji2025levitated}. In particular, the magnetic-field dependence of the hybrid eigenfrequencies $\omega_\pm$ provides a probe of the micromagnet-superconductor interaction. The resulting comb structure, with its dense set of correlated spectral lines, could enable more efficient tracking of frequency shifts induced by local imperfections of the trapping landscape or external magnetic-field perturbations. Beyond single-particle dynamics, understanding the nonlinear motion of an individual levitated micromagnet is a step toward arrays of coupled levitated magnets, in which collective modes could exhibit substantially richer nonlinear and nonequilibrium behavior~\cite{rusconi_2019_PRA}. Upon cooling to the motional quantum ground state, such levitated magnetic systems may further enable studies of macroscopic quantum phenomena in strongly nonlinear multimode regimes.

% %%%%%%%%%%%%%%%%%%%%%%%%%%%%%%%%%%%%%%%%%%%%%%%%%%
% \section{Acknowledgments}
% %%%%%%%%%%%%%%%%%%%%%%%%%%%%%%%%%%%%%%%%%%%%%%%%%%
We thank Mark Dykman, Marko Lon\v{c}ar, Uro\v{s} Deli\'{c}, Hong X. Tang, Szymon Pustelny, Shimon Kolkowitz, and Jack\,G.\,E.\,Harris for discussions and help, as well as Alex Cui and Sejoon Lim for assistance with sample magnetization.
This work was supported by the NSF Center for Ultracold Atoms, Amazon Web Services (grant No. A60290), NSF (grant No. OMA-2121044), DOE Quantum Systems Accelerator Center (grant No. DE-AC02-05CH11231), and the Air Force Office of Scientific Research (grant No. FA9550-23-1-0333). Y. W. acknowledges support from the HQI Postdoctoral Fellowship Program. T. M. acknowledges support from the NSF Graduate Research
Fellowship Program (grant No. 2140743). V.W. is supported by the Financial Support Programmes for Female Academics, from the Office for Gender Equality, Ulm University.
V.W. and B.A.S. acknowledge funding by the Carl-Zeiss-Foundation through the project QPhoton. Z.W. and B.A.S. acknowledge support from the DFG Project No. 521602913. B.A.S. is supported by the DFG No. 510794108.

% \appendix
% \onecolumngrid

%\clearpage
%\newpage
\onecolumngrid
\appendix
\def\theequation{S.\arabic{equation}}
\renewcommand{\thefigure}{S\arabic{figure}}
\setcounter{figure}{0}
\setcounter{equation}{0}

\section*{Supplementary Information for\\
Nonlinear dynamics and mechanical frequency combs with a Meissner-levitated
micromagnet}

%%%%%%%%%%%%%%%%%%%%%%%%%%%%%%%%%
\subsection{Experimental details}\label{app:exp_details}
%%%%%%%%%%%%%%%%%%%%%%%%%%%%%%%%%
The superconducting sample consists of \SI{300}{\nano\meter} thick yttrium barium copper oxide (YBCO) films grown by pulsed laser deposition on both faces of a \SI{500}{\micro\meter} thick lanthanum aluminate substrate.

	\subsubsection{Preparation of magnetic particles}
	 We use initially unmagnetized spherical particles composed of Nd-Pr-Fe-Co-Ti-Zr-B alloy~(MQP-S-11-9-20001-070 Isotropic Powder from Magnequench). To prevent the particles from agglomerating during magnetization, they are first dispersed on a silicon chip patterned with silicon nitride. A glass coverslip is then placed on top and secured using a SEM clamp. The chip is subsequently exposed to a magnetic field of \SI{8}{\tesla}, resulting in a remanent magnetic induction of approximately $B_\mathrm{rem}=\SI{0.8}{\tesla}$. This procedure leaves only particles with diameters below \SI{20}{\micro\meter} on the chip due to the competition between magnetic and surface forces. Following magnetization, the coverslip is removed and the chip is transferred into the cryostat at room temperature. Individual particles are then transferred to the superconducting surface using the \textit{in situ} probe setup described below.

	\subsubsection{Levitation procedure}
	\begin{figure}[b]
		\centering
		\includegraphics[width=\linewidth]{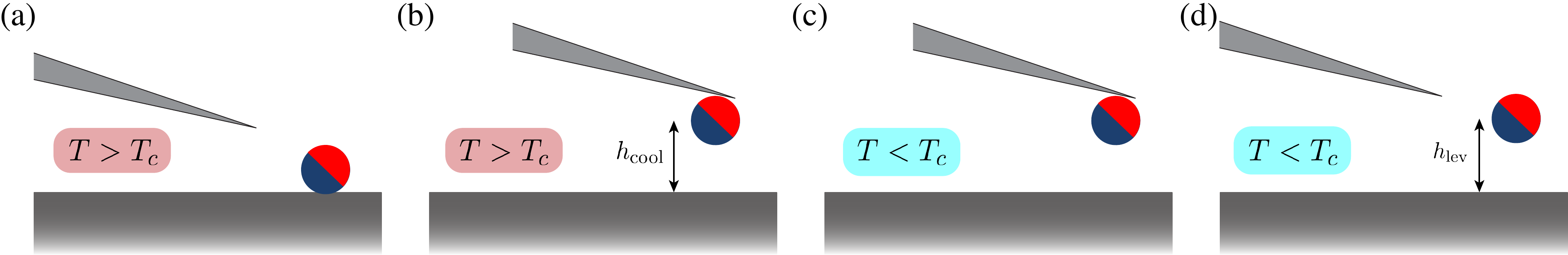}
		\caption{
		{(a)}~Above the superconducting transition temperature, $T_{\mathrm{c}}\approx\SI{87}{\kelvin}$, the micromagnet rests on the YBCO surface.
		{(b)}~A tungsten probe lifts the micromagnet to the cooldown height $h_{\mathrm{cool}}$.
		{(c)}~The cryostat is cooled below $T_{\mathrm{c}}$ to approximately \SI{6}{\kelvin}, thereby pinning the magnetic-flux distribution generated by the micromagnet. 
		{(d)}~Retraction of the probe releases the micromagnet into the magnetic trap, where it settles at the equilibrium levitation height $h_\mathrm{lev}$.}
		\label{fig:lev_proc}
	\end{figure}
	The YBCO sample is thermally anchored to the cold finger of a closed-cycle cryostat~(Montana Instruments Cryostation C2). A tungsten probe tip is mounted independently on a mass-loaded spring and positioned using a three-axis nanopositioner. The spring suppresses high-frequency mechanical vibrations of the probe relative to the sample, enabling reliable pickup of individual micromagnets (Figs.~\ref{fig:lev_proc}(a) and (b)). 
	 
	With the micromagnet held on the probe tip by surface forces, the nanopositioner is used to position the micromagnet at a cooldown height $h_{\text{cool}}$ above the superconducting surface. Up to this point, the platform temperature is kept above the superconducting transition temperature, typically at \SI{100}{\kelvin}. The platform is then cooled below the transition temperature, thereby pinning the local magnetic-flux distribution generated by the micromagnet in the type-II superconductor (Fig.~\ref{fig:lev_proc}(c)), which establishes the magnetic trapping potential~\cite{gieseler_2020_PRL, wei2026meissner}. After the temperature stabilizes at \SI{6}{\kelvin}, the probe tip is retracted from its initial position. As the tip is moved away, the magnetic restoring force exceeds the adhesion between the micromagnet and the probe, releasing the particle into the trap (Fig.~\ref{fig:lev_proc}(d)). The micromagnet subsequently settles at the equilibrium levitation height $h_\mathrm{lev}$ (center-to-surface distance), after which the probe is moved away from the measurement region.

	\subsubsection{Optical setup and readout}\label{Optical path}
\begin{figure}[t!]
			\centering
			\includegraphics[width=0.6\columnwidth]{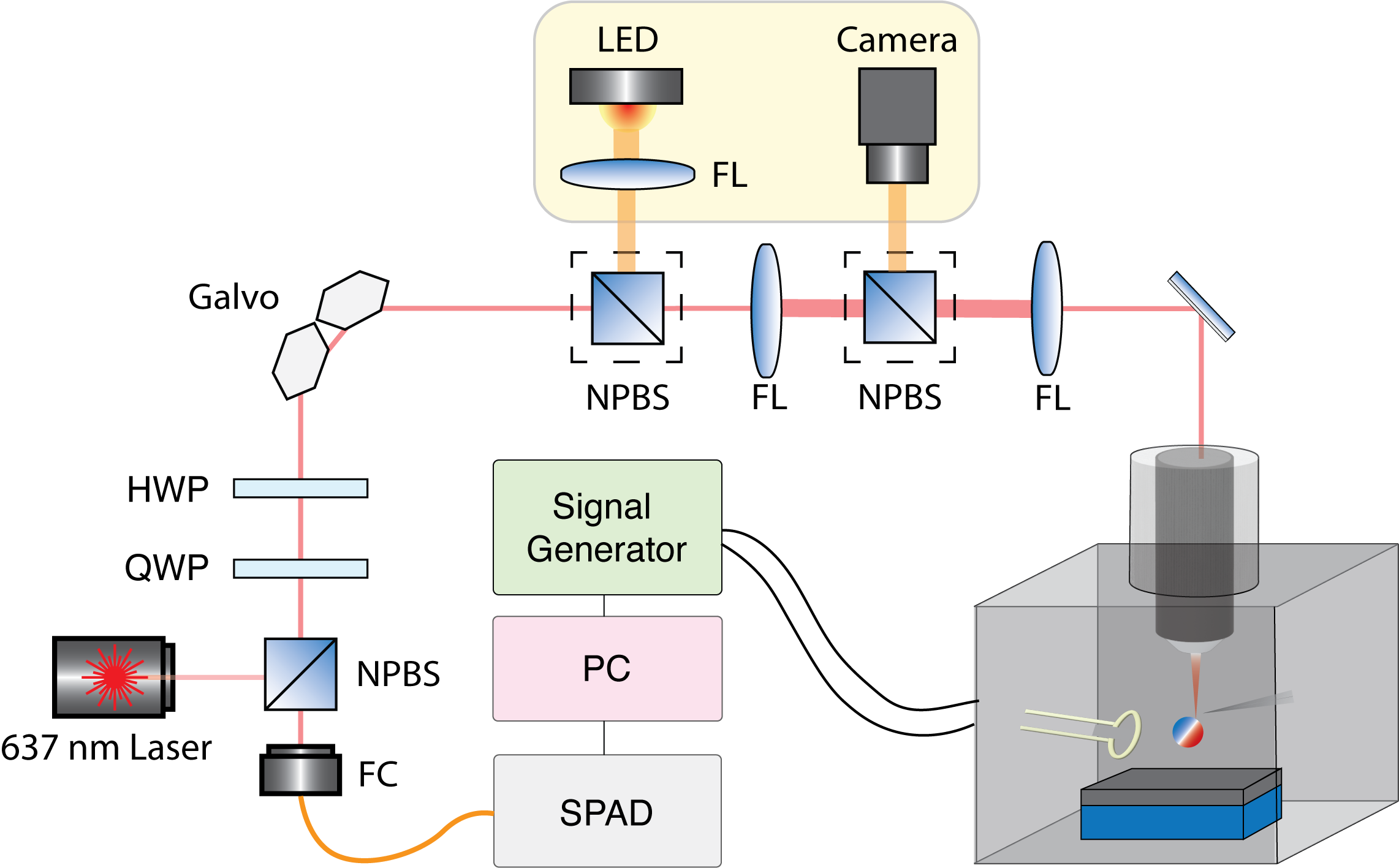}
			\caption{
			Schematic of the optical readout and \textit{in situ} micromanipulation setup. The YBCO sample and micromanipulation assembly are placed in a cryostat integrated with a 4$f$ confocal microscope. A galvanometer (Galvo) and focal lenses (FL) direct the \SI{637}{\nano\meter} readout beam through the microscope objective, enabling two-dimensional scanning of the focal point. Quarter-wave plates (QWP) and half-wave plates (HWP) control the beam polarization. Light reflected from the micromagnet is separated from the incident light by a non-polarizing beamsplitter (NPBS), coupled into an optical fiber by a fiber coupler (FC), and detected using a single-photon avalanche diode (SPAD). The LED and camera used for imaging are coupled into the beam path with flippable beamsplitters when required.}
			\label{fig:optics}
		\end{figure}

		Figure~\ref{fig:optics} shows a schematic of the optical setup used to measure the mechanical motion. We use a \SI{637}{\nano\meter} laser whose polarization is adjusted using quarter- and half-wave plates to optimize the detected photon rate. A galvanometer steers the beam through a 4$f$ confocal microscope, and the beam is focused inside the cryostat using a $100\times$ objective with a numerical aperture $\mathrm{NA}=0.8$. This configuration enables two-dimensional scanning of the laser focal point within the cryostat (see Sec.~\ref{sec:galvo_sensitivity}). The light reflected from the micromagnet is collected by the same objective and retraces the incident optical path. A non-polarizing beamsplitter (NPBS) separates the reflected light from the incident beam. The reflected light is subsequently coupled into an optical fiber and detected using a single-photon avalanche diode (SPAD), which records the arrival time of each photon. We use a camera during the levitation procedure and for coarse characterization. Broadband LED illumination and the camera imaging path are coupled into the main optical path using two flippable beamsplitters, indicated by the dashed boxes in Fig.~\ref{fig:optics}. These beamsplitters are removed during measurements.

	\subsubsection{Particle driving} \label{Particle driving}
		We use an \textit{in situ} wire bond coil inside the cryostat to generate an oscillating magnetic field that drives the mechanical modes of the levitated micromagnet. The drive signal is produced by a signal generator and subsequently amplified. Because the magnetic field shifts both the equilibrium configuration and the curvature of the trapping potential (see Sec.~\ref{sec:external_field}), it produces an additive coherent drive as well as parametric modulation of the mechanical frequencies. In addition, the mechanical modes can be driven into a thermal state by applying a frequency-modulated broadband white-noise signal.
	 
	\subsubsection{Galvanometer scans and sensitivity estimation} \label{sec:galvo_sensitivity}
\begin{figure}[t!]
			\centering
			\includegraphics[width=\columnwidth]{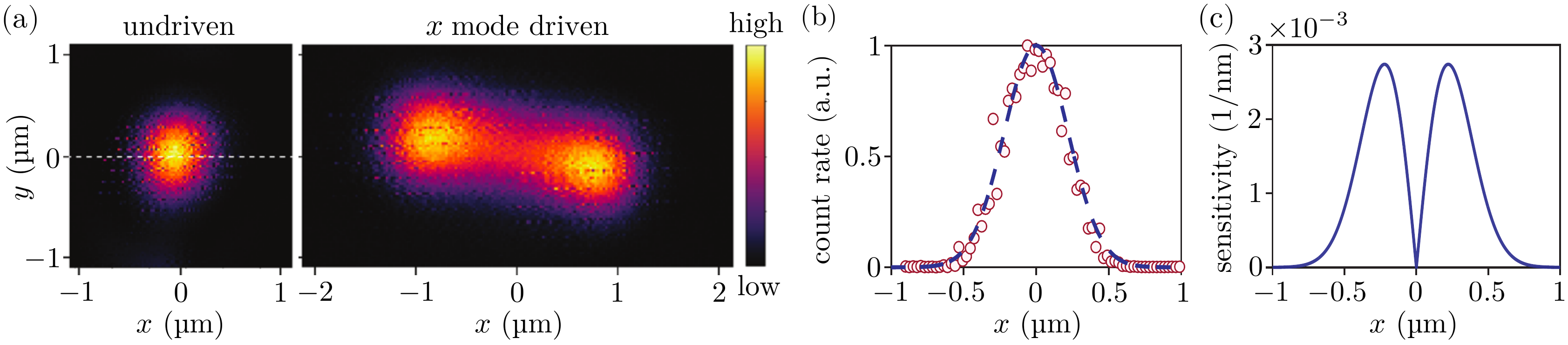}
			\caption{
			(a) Spatial scans of the micromagnet without mechanical driving (left panel) and under strong resonant coherent driving of the predominantly $x$-like hybrid mode (right panel). 
			(b)~Normalized count rate along the line cut indicated in (a). The blue dashed line is a Gaussian fit. 
			(c)~Spatial derivative of the fitted Gaussian profile, determining the displacement responsivity along the line cut.}
			\label{fig:galvo}
		\end{figure}

		We perform two-dimensional spatial scans of the optical response by rastering the focused \SI{637}{\nano\meter} readout beam across the micromagnet using the galvanometer shown in Fig.~\ref{fig:optics}. For each galvanometer setting, reflected photons are detected using the SPAD, and the corresponding photon count rate is recorded as a function of the focal-point coordinates ($x$, $y$). Figure~\ref{fig:galvo}(a) shows spatial scans of the undriven micromagnet and of the micromagnet under strong resonant driving of the predominantly $x$-like hybrid mode. 
	 
		At a fixed focal point, the motion of the micromagnet modulates the detected photon count rate. Figure~\ref{fig:galvo}(b) presents a line cut of the normalized photon count rate along the $x$-like mode direction, indicated by the white dashed line in Fig.~\ref{fig:galvo}(a). The count-rate profile is well approximated by a Gaussian function with waist $w_0$ and maximum count rate $N_0$. The count rate can therefore be modeled as
		\begin{equation}\label{eq:count_rate_gaussian}
			f_\mathrm{ph}(x) = N_0\exp \left[-\dfrac{1}{2}\left(\dfrac{x}{w_0}\right)^2\right].
		\end{equation}
		A fit to this function, shown by the blue dashed line in Fig.~\ref{fig:galvo}(b), yields a FWHM of \SI{494}{\nano\meter}, consistent with the expected diffraction-limited length scale ($\SI{486}{\nano\meter}$) for a $\lambda=\SI{637}{\nano\meter}$ beam focused by an objective with $\mathrm{NA}=0.8$. The spatial derivative of the fitted count-rate profile, shown in Fig.~\ref{fig:galvo}(c), determines the local displacement responsivity along this line cut. The maximal responsivity is obtained at $x_0=w_0$, corresponding to $2.73\times10^{-3}\,\text{nm}^{-1}$. Assuming shot-noise-limited photon detection, the corresponding displacement-noise spectral density is 
		\begin{align}\label{eq:measurement_sensitivity}
			S_{xx}=S_{NN}\left|\dfrac{\partial x}{\partial N}\right|^2={\bar{N}}\left|\dfrac{\partial x}{\partial N}\right|^2\geq\dfrac{1.4\times10^{5}\,\mathrm{nm}^2}{\bar{N}},
		\end{align}
		where $\bar{N}$ is the mean photon count rate and the shot-noise spectral density is taken to be $S_{NN}=\bar{N}$. For $\bar{N}=1\,$Mcps, corresponding to an incident laser power of approximately \SI{100}{\pico\watt} (with an improved photon-collection efficiency compared to the experiment in the Letter), we obtain a displacement sensitivity of $\sqrt{S_{xx}}\sim$\SI{0.4}{\nano\meter\per\sqrt{\hertz}}. This value is consistent with the estimated displacement sensitivity from an independent calibration. For comparison, the standard quantum limit of this mechanical mode is around \SI{0.3}{\pico\meter\per\sqrt{\hertz}}.

	\subsubsection{Frequency tuning via permanent magnet}\label{sec:freq_tuning}
		The frequencies of the hybridized in-plane mechanical modes of the levitated micromagnet can be tuned via an external magnetic field. The field couples to the magnetic moment of the micromagnet, changing its equilibrium configuration and the local curvature of the trapping potential, as described in Sec.~\ref{sec:external_field}. The field is generated by a permanent neodymium magnet mounted on a micrometer translation stage, which allows its magnitude at the micromagnet position to be varied. The magnet is aligned such that the applied field is approximately parallel to the direction of the predominantly $x$-like mode, as determined from the galvanometer scans. Translating the permanent magnet produces a field sweep spanning approximately $0.5\,$mT and centered around zero field. At each field value, we record the photon-count spectrum and extract the two hybrid-mode frequencies, $\omega_+(B_\mathrm{ext})$ and $\omega_-(B_\mathrm{ext})$. By aligning the external field with different principal directions of motion, the relative tuning rates of the modes can be controlled. The measured field dependence of $\omega_+$ and $\omega_-$ is shown in Fig.~\ref{fig1:schematic}(c), where the two frequency branches exhibit an avoided crossing characteristic of linearly coupled modes. A detailed analysis is presented in Sec.~\ref{sec:intermodal_lin_coupling}.

\subsection{Confinement and excitation of mechanical modes}\label{app:potential}
	\subsubsection{Effective trapping potential}
		To model the trapping potential of the levitated micromagnet above the type-II superconductor, we employ the frozen-image method~\cite{kordyuk1998magnetic}, which approximates the superconducting response as the sum of an instantaneous magnetic image associated with Meissner screening and a fixed frozen image representing magnetic flux pinned during field cooling. The instantaneous image follows the current position and orientation of the micromagnet, whereas the frozen image is determined by its configuration during the superconducting transition. 

		Both contributions can be expressed in terms of the magnetization distribution of the levitated magnet at center-of-mass position $\mathbf{R}=x\mathbf{e}_x+y\mathbf{e}_y+z\mathbf{e}_z$, with $z=\mathbf{R}\cdot\mathbf{e}_z$, and orientation $\Omega=(\alpha,\beta,\gamma)$, parametrized by Euler angles in the $z$-$y'$-$z''$ convention. Assuming that the particle dimensions are sufficiently small compared with the characteristic length scale over which the magnetic fields vary, we approximate the particle magnetization by a multipole expansion truncated at quadrupole order. The magnetic properties of the particle are then described by its orientation-dependent dipole moment $\mathbf{m}(\Omega)$
		and symmetric traceless quadrupole tensor $\mathrm{Q}(\Omega)$. The frozen image is centered at $\mathbf{R}_\mathrm{fr}=-z_\mathrm{fr}\mathbf{e}_z$, and is characterized by the fixed dipole moment $\mathbf{m}_\mathrm{fr}$ and quadrupole tensor $\mathrm{Q}_\mathrm{fr}$, determined by the position and orientation of the particle during field cooling~\cite{wei2026meissner}.

\begin{figure}[t!]
			\centering
			\includegraphics[width=\columnwidth]{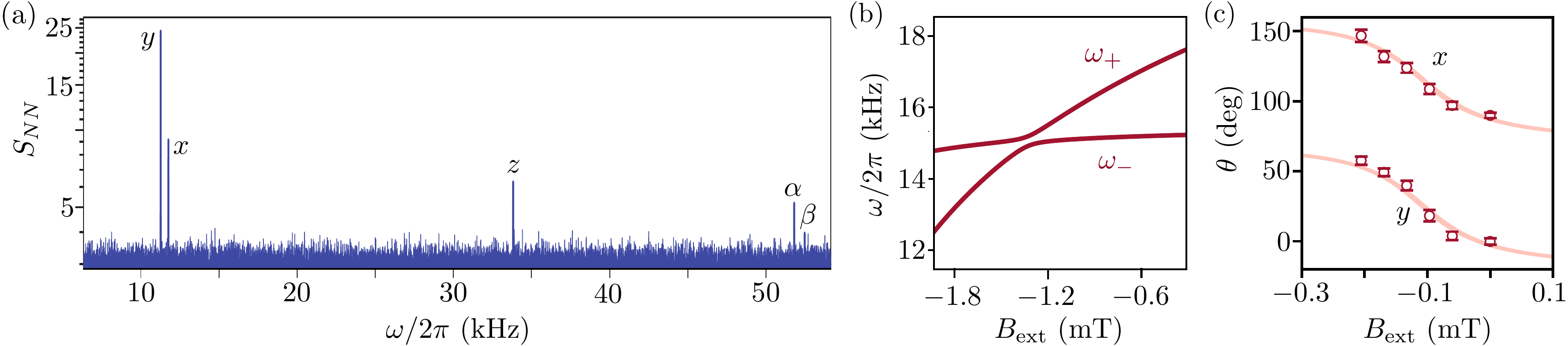}
			\caption{(a) Photon-count spectrum of the undriven micromagnet showing the mechanical resonances associated with the predominantly $y$, $x$, $z$, $\alpha$, and $\beta$ modes, ordered from low to high frequency. (b) Theoretical prediction of the hybridized in-plane translational mode frequencies $\omega_+$ and $\omega_-$ as a function of the external magnetic field $B_\mathrm{ext}$ with orientation $(\alpha_\mathrm{ext},\beta_\mathrm{ext})=(\pi/4,\pi/2)$, showing the avoided crossing associated with strong $x$-$y$ hybridization. The eigenfrequencies are obtained from Eq.~\eqref{eq:app_eigenfrequencies} evaluated at the dipole-equilibrium configuration with $\delta\beta_\mathrm{fr}=0.1$. (c) Direction of motion of the $x$- and $y$-like modes as a function of $B_\mathrm{ext}$. Solid lines show the predictions of the coupled-mode model [Eq.~\eqref{eq:mixed_angle}]. Error bars correspond to $2\sigma$ uncertainties obtained from fits to the spatial scans and estimated systematic uncertainties.}
			\label{fig:spectrum_all}
		\end{figure}

		The leading dipole contribution to the potential is
		\begin{align}\label{eqn:V_dipole}
			V_{\rm dip}({\bf R},\Omega) = &  - \dfrac{\mu_{0}}{2} \mathbf{m} \cdot \mathrm{G}(2z {\bf e}_z) \mathrm{A}\mathbf{m}  - \mu_0\mathbf{m} \cdot \mathrm{G} ( \mathbf{R} + z_{\mathrm{fr}} {\bf e}_z ) \mathbf{m}_{\mathrm{fr}},
		\end{align}
		while the leading quadrupole contribution reads
		\begin{align}\label{eqn:V_quadrupole}
			V_{\rm quad}({\bf R},\Omega)  = &   -\dfrac{\mu_0}{6} (\mathrm{Q}\nabla) \cdot \mathrm{G}(2z {\bf e}_z) \mathrm{A}\mathbf{m} - \dfrac{\mu_0}{6}\left (\mathrm{Q}\nabla \right ) \cdot \mathrm{G}(\mathbf{R} + z_{\mathrm{fr}} {\bf e}_z) \mathbf{m}_{\mathrm{fr}}  + \dfrac{\mu_0}{6} \left (\mathrm{Q}_{\mathrm{fr}} \nabla \right) \cdot \mathrm{G}(\mathbf{R} + z_{\mathrm{fr}}{\bf e}_z) \mathbf{m}.
		\end{align}
		Here $\mu_0$ denotes the vacuum permeability, ${\rm G}({\bf r}) = \nabla \otimes \nabla 1/4\pi |{\bf r}|$ is the magnetostatic Green tensor at the spatial point ${\bf r}$, and $\mathrm{A}=\mathbb{1}-2 \mathbf{e}_z\otimes\mathbf{e}_z$ is the reflection tensor associated with the planar superconducting surface with surface normal $\mathbf{e}_z$. The dipole potential $V_{\mathrm{dip}}$ is invariant under rotations of the particle about its magnetic dipole axis and therefore cannot provide a restoring torque for this degree of freedom. A quadrupole tensor that is not axially symmetric about $\mathbf{m}$ breaks this rotational symmetry and can provide confinement about the dipole axis. The quadrupole contribution therefore accounts for the finite frequency of the $\gamma$ libration. The five resonances associated with the $x$, $y$, $z$, $\alpha$, and $\beta$ modes shown in Fig.~\ref{fig:spectrum_all}(a), together with the $\gamma$ libration shown separately in Fig.~\ref{fig1:schematic}(b), demonstrate confinement of all six mechanical degrees of freedom.

		An external magnetic field $\mathbf{B}_\mathrm{ext}(\mathbf{R})$ adds the interaction
		\begin{equation}
		    V_\mathrm{ext}({\bf R},\Omega) = -\mathbf{m}\cdot\mathbf{B}_\mathrm{ext},
		\end{equation}
		where we neglect its coupling to the quadrupole moment, assuming that $\mathbf{B}_\mathrm{ext}$ varies only weakly over the dimensions of the micromagnet. The external field modifies the equilibrium position and orientation of the micromagnet and the curvature of the potential about that equilibrium, thereby tuning the frequencies and hybridization of the translational and librational modes. When the field is time dependent, the same field dependence can generate both an additive driving force and parametric modulation of the mode frequencies, as discussed below.

	\subsubsection{Tunability via external magnetic field}\label{sec:external_field}
		Neglecting spin-rotation coupling and retaining only terms up to second order in the small displacements, we describe the six mechanical degrees of freedom $q = (q_1,q_2,q_3,q_4,q_5,q_6) \equiv (x,y,z,\alpha,\beta,\gamma)$ as small oscillations about their field-dependent equilibrium configuration $q_\mathrm{eq}$~\cite{wei2026meissner}. The equilibrium is determined by $\partial_{i} V(q_\mathrm{eq})=0$, $i=1,\dots,6$, where $\partial_i=\partial/\partial q_i$. The total potential is $V=V_\mathrm{dip}+V_\mathrm{quad}+V_\mathrm{ext}$, where we omit gravity since it produces a negligible shift of the equilibrium configuration under the present experimental conditions. Expanding about $q_\mathrm{eq}$, with $\delta q = q-q_\mathrm{eq}$, gives 
		\begin{equation}
			V_\mathrm{harm}\approx V(q_\mathrm{eq}) + \dfrac{1}{2} \sum_{i,j=1}^6 K_{ij}\, \delta q_i \delta q_j,
			\label{eq:V_harmonic_expansion}
		\end{equation}
		where the field-dependent stiffness matrix is defined as
		\begin{equation}
			K_{ij}= \partial_i \partial_j V(q_\mathrm{eq}).
			\label{eq:linear_coupling_coefficients}
		\end{equation} 
		The linearized equations of motion follow from the harmonic Hamiltonian~\cite{wei2026meissner} as
		\begin{equation}
			\sum_{j=1}^6\left( M_{ij}\,\delta\ddot{q}_j+K_{ij}\,\delta q_j\right)=0,
			\label{eq:eom_harmonic_potential}
		\end{equation}
		where $\mathrm{M}$ is the generalized mass matrix evaluated at the equilibrium configuration. In a suitable local basis of Cartesian displacements and infinitesimal rotations, $\mathrm{M}$ can be taken to be diagonal, with entries given by the particle mass $M$ and the appropriate moments of inertia $I_1, I_2, I_3$. The mechanical eigenfrequencies follow from 
		\begin{equation}\label{eq:app_eigenfrequencies}
			\mathrm{det}\!\left(\mathrm{K}-\omega^2\mathrm{M}\right)=0.
		\end{equation}
		The external magnetic field affects the dynamics in two ways: (i) by shifting the equilibrium configuration $q_\mathrm{eq}$ and (ii) by modifying the curvature matrix $K_{ij}$, thereby changing both the eigenfrequencies and the hybridization of the mechanical modes. 

		To illustrate the physical origin of this tunability, consider a spatially homogeneous field whose direction is specified by the angles $\alpha_\mathrm{ext}$ and $ \beta_\mathrm{ext}$. Assuming that the magnetic dipole moment is aligned with the body-fixed axis $\mathbf{n}_3=\cos\alpha\sin\beta\,\mathbf{e}_x+\sin\alpha\sin\beta\,\mathbf{e}_y+\cos\beta\,\mathbf{e}_z$, the Zeeman potential is
		\begin{align}
			V_\mathrm{ext}=&-|\mathbf{m}|B_\mathrm{ext}\left[\cos\beta\cos\beta_\mathrm{ext}+\cos(\alpha-\alpha_\mathrm{ext})\sin\beta\sin\beta_\mathrm{ext}\right].
		\end{align}
		Because a homogeneous field has no direct dependence on the center-of-mass coordinates, it acts directly on the rotational degrees of freedom. Its effect is transferred to the translational motion through the rotation–translation couplings in the trapping potential. To obtain a transparent reduced model for a field applied along the $x$ direction, we retain the dominant coupling between the translational coordinate $x$ and the rotational coordinate $\beta$. We expand about the zero-bias equilibrium of the full trapping potential. Its orientation is dominated by the dipole interaction with the frozen image and is characterized by $\alpha_\mathrm{fr}=0$ and $\beta_\mathrm{fr}=\pi/2$~\cite{wei2026meissner}. Considering a small frozen-image tilt $\delta\beta_\mathrm{fr}\ll 1$ out of the $x$-$y$ plane results in, to leading order, the generalized torque
		\begin{equation}\label{eq:app_torque_beta}
			N_\beta = -\partial_\beta V_\mathrm{ext}\approx -|\mathbf{m}|B_\mathrm{ext}\delta\beta_\mathrm{fr}.
		\end{equation}
		Retaining only the $x$-$\beta$ subsystem, the reduced equations of motion follow from Eqs.~\eqref{eq:eom_harmonic_potential} and \eqref{eq:app_torque_beta} as
		\begin{subequations}
			\begin{align}
				\delta \ddot{x}& = -\dfrac{K_{xx}}{M}  \delta x - \dfrac{K_{x\beta} }{M} \delta\beta,\\
				\delta \ddot{\beta} &= -\dfrac{ K_{\beta\beta}(B_\mathrm{ext})}{I_1} \delta \beta - \dfrac{K_{x\beta }}{I_1}  \delta x - \dfrac{|\mathbf{m}|B_\mathrm{ext}}{I_1}\delta\beta_\mathrm{fr},
			\end{align}
		\end{subequations}
		where the rotational stiffness component $K_{\beta\beta}(B_\mathrm{ext})$ is field-dependent and $K_{x\beta}=K_{\beta x}$. If the rotational mode frequency $\omega_\beta=\sqrt{K_{\beta\beta}(B_\mathrm{ext})/I_1}$ is large compared with both the translational dynamics and the applied field modulation frequency, the rotational coordinate can be adiabatically eliminated by setting $\delta\ddot{\beta}\approx 0$. This gives
		\begin{equation}\label{eq:app_ddotx}
			\delta\ddot{x} = -\left[\dfrac{K_{xx}}{M} - \dfrac{K_{x\beta}^2}{M K_{\beta\beta}(B_\mathrm{ext})}\right] \delta x + \dfrac{|\mathbf{m}|B_\mathrm{ext} K_{x\beta}}{M K_{\beta\beta}(B_\mathrm{ext})}\delta\beta_\mathrm{fr}.
		\end{equation}
		The effective uncoupled $x$-coordinate frequency, after accounting for the coupling to $\beta$ but before including $x$-$y$ hybridization, is therefore
		\begin{equation}\label{eq:app_frequency_x}
			\omega_x^2(B_\mathrm{ext})=\dfrac{K_{xx}}{M}- \dfrac{K_{x\beta}^2}{M K_{\beta\beta}(B_\mathrm{ext})}.
		\end{equation} 
		Equation~\eqref{eq:app_ddotx} shows that a homogeneous magnetic field induces both a frequency shift and, in the presence of a finite misalignment $\delta\beta_\mathrm{fr}$ between the applied field and the frozen dipole orientation, a shift of the equilibrium position. While the derivation above assumes a finite frozen-image tilt $\delta\beta_\mathrm{fr}$, an equivalent effect arises if the external magnetic field is tilted out of plane instead. 

		In the experiment, the static bias field is supplied by the permanent magnet, whereas the \textit{in situ} coil produces a time-dependent component $B_\mathrm{ext}(t)\propto\cos(\omega_\mathrm{d}t)$ with drive frequency $\omega_\mathrm{d}$. The oscillating part provides an additive coherent drive. At the same time, the field dependence of $K_{\beta\beta}$, and hence $\omega_x$, produces parametric modulation of the translational frequency. To leading order, the resulting equation has the form
		\begin{equation}
			\delta\ddot{x} +\left[\omega_x^2-\varepsilon\cos(\omega_\mathrm{d}t)\right] \delta x = \eta \cos(\omega_\mathrm{d}t),
		\end{equation}
		where $\varepsilon$ and $\eta$ describe the parametric and additive drive strengths, respectively.

		An analogous mechanism applies to the $y$ coordinate through its coupling to the corresponding rotational degree of freedom $\alpha$. Thus, a magnetic field applied parallel to the superconducting surface provides a direct means to tune the effective coordinate frequencies, while its time-dependent component can produce both coherent forcing and parametric modulation, as discussed in Sec.~\ref{sec:freq_tuning}.
	 
	\subsubsection{Intermodal linear coupling}\label{sec:intermodal_lin_coupling}
		The external magnetic field modifies the eigenfrequencies of the mechanical modes. Because the translational and rotational degrees of freedom are generally coupled, the mechanical eigenmodes are hybridized~\cite{wei2026meissner}; labels such as $x$-like or $y$-like therefore indicate the dominant direction of motion far from the avoided crossing and do not imply purely translational eigenmodes. Experimentally, we apply external magnetic fields up to \SI{300}{\micro\tesla} approximately parallel to the predominantly $x$-like direction and sweep through the resonance between the two in-plane modes. As shown in Fig.~\ref{fig1:schematic}(c), the measured eigenfrequencies $\omega_+(B_\mathrm{ext})$ and $\omega_-(B_\mathrm{ext})$ exhibit an avoided crossing with minimum splitting near \SI{-110}{\micro\tesla}, indicating strong linear intermodal coupling between the underlying in-plane degrees of freedom.

		To describe this behavior, we consider the effective $x$-$y$ subsystem obtained after eliminating the fast rotational degree of freedom discussed in Sec.~\ref{sec:external_field}. Diagonalizing the resulting linear coupled-mode equations yields the hybridized eigenfrequencies 
		\begin{equation}\label{eq:coupled_mode_freq}
			\omega_{\pm}^2(B_\mathrm{ext})=\dfrac{\omega_{x}^2+\omega_{y}^2}{2}\pm\sqrt{\dfrac{(\omega_{x}^2-\omega_{y}^2)^2}{4}+k_{xy}^2},
		\end{equation}
		where $\omega_{x}(B_\mathrm{ext})$ and $\omega_{y}(B_\mathrm{ext})$ denote the effective uncoupled translational frequencies and $k_{xy} = K_{xy}/M$ is the linear intermodal coupling coefficient. At resonance, $\omega_0\approx\omega_{x}\approx\omega_{y}$, the eigenfrequencies reduce to $\omega_\pm\approx\omega_0\pm k_{xy}/2\omega_0$, yielding a minimum splitting $\Delta \omega_\mathrm{min}\approx k_{xy}/\omega_0=2g$. This provides a direct measure of the effective coupling rate $g=k_{xy}/2\omega_0$. The measured eigenfrequencies are fitted to Eq.~\eqref{eq:coupled_mode_freq} in Fig.~\ref{fig1:schematic}(c), assuming smooth field dependences of the effective uncoupled frequencies $\omega_{x}(B_\mathrm{ext})$ and $\omega_{y}(B_\mathrm{ext})$, yielding a coupling rate $g/2\pi=\SI{511\pm10}{\hertz}$. The theoretical curves in Fig.~\ref{fig:spectrum_all}(b) are obtained by evaluating Eq.~\eqref{eq:app_eigenfrequencies} at the dipole-equilibrium configuration, with an out-of-plane frozen-image tilt of $\delta\beta_\mathrm{fr}=0.1$ and an external field orientation $(\alpha_\mathrm{ext},\beta_\mathrm{ext})=(\pi/4,\pi/2)$. The resulting curves reproduce the measured field dependence qualitatively. 

		The hybridization additionally leads to a continuous rotation of the mode axes. The mixing angle $\theta$, defined by the relative contribution of the $x$ and $y$ motion to the hybridized eigenmodes, is given by
		\begin{align}\label{eq:mixed_angle}
			\theta(B_\mathrm{ext})=\dfrac{1}{2}\arctan\left(\dfrac{2k_{xy}}{\omega_{x}^2-\omega_{y}^2}\right).
		\end{align}
		Near resonance $\omega_0$, the relation reduces to $\tan(2\theta)\approx 2g/(\omega_{x}-\omega_{y})$. The mixing angle is extracted experimentally from galvanometer scans (Sec.~\ref{sec:galvo_sensitivity}) by measuring the directions of motion of the modes while sweeping $B_\mathrm{ext}$ through the avoided crossing, and shown in Fig.~\ref{fig:spectrum_all}(c). The measured mode directions remain orthogonal while rotating continuously across the resonance, in agreement with the coupled-mode model, confirming coherent hybridization of the two mechanical modes. The shaded region indicates a $2\sigma$ confidence interval obtained from the fit to Eq.~\eqref{eq:coupled_mode_freq}, including estimated systematic uncertainties.

	\subsubsection{Expansion beyond the linear regime}\label{sec:nonlinear_expansion}
		Under strong periodic driving, higher-order terms in the trapping potential become relevant, giving rise to amplitude-dependent frequency shifts and nonlinear intermodal coupling. We expand the potential about the field-dependent equilibrium configuration to fourth order, such that Eq.~\eqref{eq:V_harmonic_expansion} is extended to $V\approx V_\mathrm{harm} + V_\mathrm{nl}$. Projecting onto the in-plane translational coordinates $x$ and $y$ and the $\gamma$ libration, while incorporating the effects of the remaining coordinates into the effective mode parameters and magnetic-drive terms [Sec.~\ref{sec:external_field}], and retaining only the terms relevant for the minimal model in Eqs.~\eqref{eq:eom_xy} and~\eqref{eq:eom_gamma}, the nonlinear part of the reduced potential is
		\begin{align}
			V_\mathrm{nl}=\dfrac{M\lambda_x}{4}x^4+\dfrac{M\lambda_y}{4}y^4+\dfrac{I_\gamma\lambda_\gamma}{4}\gamma^4+\dfrac{M g_2}{2}x^2y^2+I_\gamma\chi x\gamma^2+I_\gamma\zeta x^2\gamma+\cdots.
			\label{eq:V_nl_reduced}
		\end{align}
		Here, $I_\gamma$ is the moment of inertia associated with the $\gamma$ libration. The coefficients $\lambda_i$, with $i=x,y,\gamma$, describe the self-Duffing nonlinearities, $g_2$ describes the quartic coupling between the translational coordinates, and $\chi$ and $\zeta$ describe the leading cubic interactions between $x$ and $\gamma$. They are defined as
		\begin{equation}
			\lambda_i=\dfrac{1}{6 M_{ii}}\partial^4_i V(q_\mathrm{eq}),\qquad g_2=\dfrac{1}{2M}\partial_x^2\partial_y^2V(q_\mathrm{eq}),\qquad\chi=\dfrac{1}{2I_\gamma}\partial_x\partial_\gamma^2V(q_\mathrm{eq}),\qquad \zeta=\dfrac{1}{2I_\gamma}\partial^2_x\partial_\gamma V(q_\mathrm{eq}),
			\label{eq:nonlinear_coefficient_definitions}
		\end{equation}
		with $M_{xx}=M_{yy}=M$ and $M_{\gamma\gamma}=I_\gamma$. Because the relevant mechanical eigenmodes contain substantial admixtures of the remaining translational and librational degrees of freedom, assigning independent numerical values to the coefficients of a strictly three-coordinate reduction would depend on the chosen projection. We therefore do not quote these coefficients individually. Nevertheless, direct evaluation of the derivatives of the full trapping potential confirms that the nonlinear interactions retained in the reduced model are nonzero. Within this projection, the $x\gamma^2$ term proportional to $\chi$ is the dominant cubic intermodal interaction among all couplings involving $x$, $y$, and $\gamma$.

\subsection{Characterization of nonlinear micromagnet dynamics}\label{app:characterizing_nonlinearities}
	We characterize the mechanical damping, driven nonlinear response, and nonlinear intermodal coupling of the levitated micromagnet through ringdown and driven-response measurements.
	 
	\subsubsection{Quality factors}\label{sec:ringdown}
\begin{figure}[t!]
			\centering
			\includegraphics[width=\columnwidth]{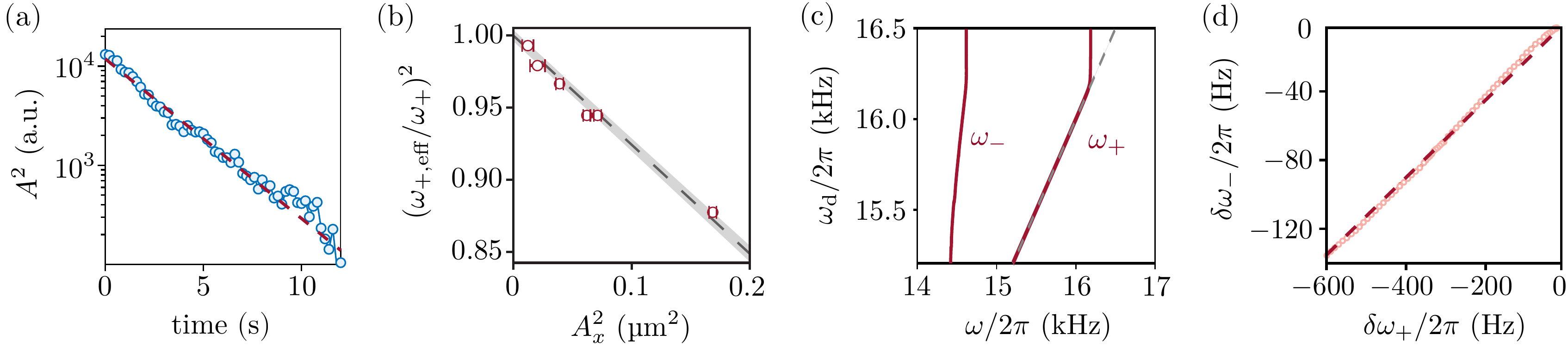}
			\caption{
			(a) Measured energy of the predominantly $x$-like mode, inferred from the spectral area, as a function of time during repeated ringdown measurements. The red dashed line shows a fit to the exponential decay. 
			(b) Normalized squared effective frequency $(\omega_{+,\text{eff}}/\omega_+)^2$ as a function of the squared $x$-projected oscillation amplitude $A^2_x$. Error bars indicate $2\sigma$ uncertainties obtained from fits to spatial scans and estimated systematic uncertainties. The gray dashed line shows a linear fit, and the light gray shaded region denotes its 95\% confidence interval. 
			\textrm{(c)}~Spectral response of the $y$-like mode with frequency $\omega_-$ while the $x$-like mode with frequency $\omega_+$ is coherently driven during a downward sweep of $\omega_{\mathrm{d}}$ (gray dashed line).
			\textrm{(d)}~Measured lower-mode frequency shift $\delta \omega_-$ as a function of the corresponding upper-mode shift $\delta \omega_+$. The dashed line shows a linear fit.}
			\label{fig:ringdown}
		\end{figure}

		We determine the quality factors $Q_\pm$ of the hybrid mechanical modes from ringdown measurements. Each mode is first excited resonantly using the \textit{in situ} drive coil. The resulting increase in the integrated spectral area of the corresponding resonance is proportional to the mode energy and therefore to the squared oscillation amplitude $A_i^2$. After the drive is switched off, the micromagnet undergoes free decay, during which the mode energy decreases exponentially due to dissipation. To monitor this decay, we record photon-count spectra in \SI{0.2}{\second} intervals and extract the spectral area associated with the mode of interest in each interval. Figure~\ref{fig:ringdown}(a) shows the measured ringdown of the predominantly $x$-like upper hybrid mode obtained from repeated measurements. The data are fitted with
		\begin{equation}
			A_x^2(t) = A_x^2(0)\,\mathrm{e}^{-t/\tau_0},
		\end{equation}
		yielding a decay time constant of $\tau_0=\SI{2.85(8)}{\second}$. Identifying the energy decay rate with the mechanical damping rate, $\Gamma_+=1/\tau_0$, gives a quality factor of $Q_+=\omega_+/\Gamma_+=2.95(8)\times10^5$. Applying the same procedure to the predominantly $y$-like lower hybrid mode yields $Q_-=2.0(3)\times10^5$. Across measurements on this platform, the mechanical quality factors typically range from $10^5$ to $10^6$.

	\subsubsection{Driven nonlinear mechanical response }\label{sec:nonlinear_response}
		To characterize the Duffing nonlinearity through the response shown in Fig.~\ref{fig1:schematic}(d), we first consider the single-mode response of the $x$ coordinate, neglecting intermodal coupling and the nonresonant parametric-drive contribution,
		\begin{equation}
			\ddot{x}+\Gamma_x\dot{x}+\omega_x^2 x +\lambda_x x^3=\eta\cos(\omega_\mathrm{d}t).
			\label{eq:eom_x_Duffing_driven}
		\end{equation}
		For near-resonant driving, $\omega_\mathrm{d}\approx\omega_x$, and in the weak-damping regime $\Gamma_x\ll \omega_x$ relevant for the high-$Q$ mechanical modes, the steady-state motion is well approximated by the single-harmonic ansatz
		\begin{equation}
			x(t)=A_x \cos(\omega_\mathrm{d}t-\phi),
			\label{eq:x_harmonic_ansatz}
		\end{equation}
		where $A_x$ and $\phi$ denote the steady-state oscillation amplitude and phase, respectively. Substituting this ansatz into Eq.~\eqref{eq:eom_x_Duffing_driven}, retaining only the resonant contribution of the cubic nonlinearity within a rotating-wave approximation (RWA), $\cos^3(\omega_\mathrm{d}t-\phi)\rightarrow 3\cos(\omega_\mathrm{d}t-\phi)/4$, and equating the coefficients of $\cos(\omega_\mathrm{d}t)$ and $\sin(\omega_\mathrm{d}t)$, we eliminate $\phi$ by squaring and adding the resulting equations. This yields
		\begin{equation}
			\eta^2=A_x^2\left[\left(\omega_x^2-\omega_\mathrm{d}^2+\dfrac{3\lambda_x A_x^2}{4}\right)^2+(\Gamma_x\omega_\mathrm{d})^2\right].
		\end{equation}
		Introducing the detuning $\Delta_x=\omega_\mathrm{d}-\omega_x$ with $|\Delta_x|\ll\omega_x$, and using $\omega_x^2-\omega_\mathrm{d}^2\approx-2\omega_x\Delta_x$, the above steady-state amplitude equation becomes
		\begin{align}\label{eq:duffing_amp}
			A_x^2=\dfrac{\eta^2 / 4 \omega_x^2}{\left(\Delta_x -3\lambda_x A_x^2/8 \omega_x\right)^2+\Gamma_x^2/4}.
		\end{align}
		The Duffing nonlinearity therefore produces an amplitude-dependent frequency shift
		\begin{equation}\label{eq:app_duffing_shift}
			\delta\omega_\mathrm{duff}=\dfrac{3\lambda_x A_x^2}{8 \omega_x},
		\end{equation}
		or, equivalently, defines the amplitude-dependent effective resonance frequency 
		\begin{equation}
			\omega_{x,\mathrm{eff}}^2=\omega_{x}^2+\dfrac{3\lambda_x A_x^2}{4}.
		\end{equation}
		For sufficiently strong drive, Eq.~\eqref{eq:duffing_amp} admits multiple steady-state solutions, giving rise to the familiar bistability of a driven Duffing oscillator. This behavior is observed experimentally in the driven response of the predominantly $x$-like mode shown in Fig.~\ref{fig1:schematic}(d). As the drive frequency $\omega_{\mathrm{d}}$ is swept across resonance, the oscillation amplitude exhibits hysteresis between upward and downward frequency sweeps. The discontinuous amplitude jumps mark transitions between the stable Duffing branches~\cite{huber_2020_PRX,ochs_2022_PRX}. The observed bending toward lower frequencies indicates a negative Duffing nonlinearity. 

		To apply the single-mode result to the measured linearly hybridized mode, we make the replacements $\omega_x\rightarrow\omega_+$, $\omega_{x,\mathrm{eff}}\rightarrow\omega_{+,\mathrm{eff}}$, $\Gamma_x\rightarrow\Gamma_+$, and $\Delta_x\rightarrow\Delta_+$. From the measured dependence of $\omega_{+,\mathrm{eff}}$ on $A_x$ [Fig.~\ref{fig:ringdown}(b)], we extract the effective Duffing coefficient $\lambda_x/(2\pi)^2=-1.5(3)\times 10^{22}~\mathrm{Hz^2/m^2}$. Analogously, applying the same analysis for the $y$-like mode yields $\lambda_y/(2\pi)^2=-0.5(1)\times 10^{22}~\mathrm{Hz^2/m^2}$.

	\subsubsection{Thermal-noise-induced sidebands}\label{sec:thermal_noise_sidebands}
		In addition to the deterministic driven response, thermal fluctuations around the steady state give rise to characteristic features in the frequency spectrum. In particular, the spectrum of a driven Duffing oscillator exhibits a coherent peak at the drive frequency $\omega_\mathrm{d}$, accompanied by two noise-induced sidebands located symmetrically about $\omega_\mathrm{d}$. To determine the sideband frequencies, we express the noisy Duffing dynamics in terms of a slowly varying complex amplitude $a(t)$ in a frame rotating at the drive frequency,
		\begin{equation}
			x(t)=\dfrac{1}{\sqrt{2}}\left[a(t)\mathrm{e}^{-i\omega_\mathrm{d}t}+a^\ast(t)\mathrm{e}^{i\omega_\mathrm{d}t}\right].
		\end{equation}
		Assuming a slowly varying envelope, $|\dot{a}|\ll\omega_\mathrm{d}|a|$, we apply an RWA. Expanding the resulting equation of motion about the stationary driven solution $a_\mathrm{s}$, $a(t)=a_\mathrm{s}+\delta a(t)$, yields the linearized dynamics of the noise-driven fluctuations $\delta a$,
		\begin{equation}
			\delta\dot{a}=-\dfrac{\Gamma_x}{2}\delta a+i\left(\Delta_x-\dfrac{3\lambda_x|a_\mathrm{s}|^2}{2\omega_\mathrm{d}}\right)\delta a - i\dfrac{3\lambda_x a_\mathrm{s}^2}{4\omega_\mathrm{d}}\delta a^\ast+\sqrt{\dfrac{\Gamma_x k_\mathrm{B}T}{M\omega_\mathrm{d}^2}} \xi_\mathrm{th}(t).
		\end{equation}
		Here, $k_\mathrm{B}$ is the Boltzmann constant, and $\xi_\mathrm{th}(t)$ is a normalized, zero-mean complex Gaussian white-noise process representing the resonant thermal fluctuations in the rotating frame. Its second moments satisfy $\mathbb{E}[\xi_\mathrm{th}(t)\xi^\ast_\mathrm{th}(t')]=\delta(t-t')$ and $\mathbb{E}[\xi_\mathrm{th}(t)\xi_\mathrm{th}(t')]=0$. The coupling between $\delta a$ and $\delta a^\ast$ is a direct consequence of the nonlinear driven steady state. It produces oscillatory fluctuations in the rotating frame, which appear as sidebands in the laboratory-frame spectrum. 

		The positive frequency offset $\delta\omega_\mathrm{s}\geq 0$ of either sideband from the drive, $\omega_\mathrm{d}\pm\delta\omega_\mathrm{s}$, is determined by the eigenvalues of the linearized dynamics. Neglecting the noise term and writing the coupled equations for $\delta a$ and $\delta a^\ast$ in matrix form yields the eigenvalues
		\begin{equation}
			s_\pm = -\dfrac{\Gamma_x}{2}\pm i\delta\omega_\mathrm{s},
		\end{equation}
		where
		\begin{equation}
			\delta\omega_\mathrm{s}^2=\left(\Delta_x-\dfrac{3\lambda_x|a_\mathrm{s}|^2}{2\omega_\mathrm{d}}\right)^2-\left(\dfrac{3\lambda_x|a_\mathrm{s}|^2}{4\omega_\mathrm{d}}\right)^2.
		\end{equation}
		Using Eq.~\eqref{eq:app_duffing_shift}, together with $A_x=\sqrt{2}|a_\mathrm{s}|$ and the near-resonant approximation $\omega_\mathrm{d}\approx\omega_x$, we obtain
		\begin{equation}\label{eq:sat_freq}
			\delta\omega_{\mathrm{s}}=\sqrt{(\delta\omega_\mathrm{duff}-\Delta_x)(3\delta\omega_\mathrm{duff}-\Delta_x)}.
		\end{equation}

\begin{figure}[t!]
			\centering
			\includegraphics[width=0.75\columnwidth]{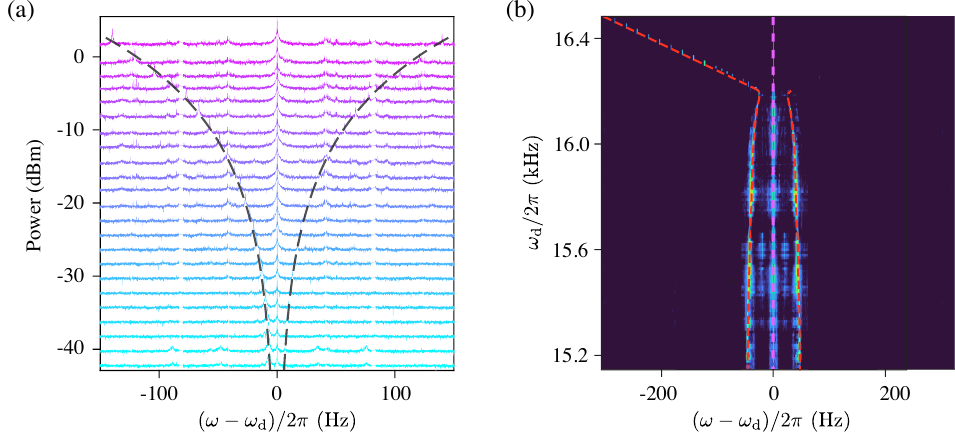}
			\caption{
			(a) Spectrum of the coherently driven $x$-like mode for several drive powers. Frequency is shown relative to the drive frequency $\omega_{\mathrm{d}}$. The black dashed lines show the $P^{1/3}$ scaling predicted by Eq.~\eqref{eq:sat_freq_vs_power}. 
			(b) Spectrum of the $x$-projected motion during a downward sweep of $\omega_{\mathrm{d}}$, with frequency again shown relative to the drive. The purple dashed line marks the drive tone, while the red dashed lines indicate two thermal-noise-induced sidebands.}
			\label{fig:sat_peak_driven_nonlinear}
		\end{figure}

		In the regime $|\Delta_x|\ll|\delta\omega_\mathrm{duff}|$, Eq.~\eqref{eq:sat_freq} reduces to $\delta\omega_{\mathrm{s}}\approx\sqrt{3}|\delta\omega_\mathrm{duff}|$. Using Eq.~\eqref{eq:duffing_amp}, the magnitude of the Duffing shift scales with the drive power $P$ as $|\delta\omega_\mathrm{duff}|\propto P^{1/3}$ in the strongly nonlinear regime, and hence
		\begin{align}\label{eq:sat_freq_vs_power}
			\delta\omega_{\mathrm{s}}\propto P^{1/3}.
		\end{align}
		In the opposite limit, $|\Delta_x|\gg|\delta\omega_\mathrm{duff}|$, Eq.~\eqref{eq:sat_freq} gives $\delta\omega_{\mathrm{s}}\approx|\Delta_x|$. Figure~\ref{fig:sat_peak_driven_nonlinear}(a) shows spectra recorded at resonance for several drive powers, where two sidebands are observed symmetrically about the drive frequency $\omega_{\mathrm{d}}$. The measured sideband offsets follow the scaling predicted by Eq.~\eqref{eq:sat_freq_vs_power}, as indicated by the dashed lines. Equation~\eqref{eq:sat_freq} therefore provides experimental access to the Duffing-induced frequency shift and, through Eq.~\eqref{eq:app_duffing_shift}, the oscillation amplitude.

		In the experiment, the drive frequency is swept across the resonance at fixed drive power, starting above (below) the undriven resonance frequency and sweeping downward (upward). Figure~\ref{fig:sat_peak_driven_nonlinear}(b) shows the measured spectrum during a downward frequency sweep, centered at the drive frequency (purple dashed line). At large positive detuning, $\omega_{\mathrm{d}}>\omega_+=2\pi\times\SI{16.44}{\kilo\hertz}$, the response is weak and the sidebands are not resolved. As the drive approaches resonance, two sidebands emerge symmetrically about the drive tone, as highlighted by the red dashed lines. Their frequency separation evolves continuously throughout the sweep, reflecting the changing oscillation amplitude of the nonlinear steady state. 

		For the experimental $x$-like hybrid mode, Eq.~\eqref{eq:sat_freq} is evaluated using the substitution $\Gamma_x\rightarrow\Gamma_+$, $\Delta_x\rightarrow\Delta_+$ introduced in Sec.~\ref{sec:nonlinear_response}. Using the independently determined Duffing coefficient $\lambda_x$ [Fig.~\ref{fig:ringdown}(b)], the measured values of $\delta\omega_\mathrm{s}$ can be converted into $A_x$. In the experimentally relevant regime, $|\Delta_+-\delta\omega_\mathrm{duff}|\ll |\delta\omega_\mathrm{duff}|$, Eq.~\eqref{eq:sat_freq} reduces, for the measured softening nonlinearity $\delta\omega_\mathrm{duff}<0$, to
		\begin{equation} 
			\delta\omega_{\mathrm{s}}\approx\sqrt{2|\delta\omega_\mathrm{duff}|(\Delta_+-\delta\omega_\mathrm{duff})}.
		\end{equation}
		The amplitudes extracted from the sideband offsets reproduce the characteristic Duffing response shown in Fig.~\ref{fig1:schematic}(d). In particular, the branches obtained during downward and upward frequency sweeps correspond to the high- and low-amplitude stable solutions of Eq.~\eqref{eq:duffing_amp}, respectively. 

	\subsubsection{Nonlinear coupling}\label{app:nonlinear_coupling}
		In addition to the intrinsic Duffing nonlinearities, 
		the trapping potential contains higher-order intermodal interactions, such as terms proportional to $x^2 y^2$, whose effects we observe experimentally.
		This coupling primarily produces amplitude-dependent frequency renormalization and mode pulling~\cite{chen_2017_NatComms, westra_2010_PRL, eichler_2012_PRL, gieseler_2014_PRL}, whereby the resonance frequency associated with one coordinate shifts with the oscillation amplitude of the other. The leading contribution retained in the reduced potential, Eq.~\eqref{eq:V_nl_reduced}, is $M g_2 x^2 y^2/2$.

		To determine the frequency pulling associated with this interaction, we treat the strongly driven $x$ motion as a large coherent oscillation, while assuming that the $y$ motion remains weakly excited. We therefore neglect the $y$-mode Duffing nonlinearity $\lambda_y$ and the backaction of $y$ on $x$. Substituting the harmonic response from Eq.~\eqref{eq:x_harmonic_ansatz} into the resulting equation of motion for $y$ yields
		\begin{equation}
			\ddot{y}+\Gamma_y\dot{y}+\left[\omega_y^2+\dfrac{g_2 A_x^2 }{2}\left[1+\cos(2\omega_\mathrm{d} t-2\phi)\right]\right]y=0.
		\end{equation}
		The nonlinear interaction therefore produces both a static frequency renormalization and a parametric modulation of the $y$-mode frequency at $2\omega_\mathrm{d}$. Away from parametric resonances of the $y$ motion, the oscillatory contribution is nonresonant at leading order, and the frequency shift is governed by the time-averaged mean-square displacement $\langle x^2\rangle=A_x^2/2$. The resulting effective resonance frequency is therefore
		\begin{equation}
			\omega_{y,\mathrm{eff}}^2=\omega_y^2+\dfrac{g_2A_x^2}{2}.
		\end{equation}
		For small frequency shifts $\delta\omega_y=\omega_{y,\mathrm{eff}}-\omega_{y}$ with $|\delta\omega_y|\ll \omega_{y}$, the effective resonance frequency can be expanded to first order, yielding $ \delta\omega_y\approx g_2 A_x^2/4\omega_y$. Comparing this expression with the Duffing shift of the driven $x$ coordinate [Eq.~\eqref{eq:app_duffing_shift}] gives
		\begin{equation}\label{eq:pulling_linear_relation}
			\dfrac{\delta\omega_y}{\delta\omega_\mathrm{duff}}=\dfrac{2 g_2 \omega_x}{3\lambda_x \omega_y}.
		\end{equation}
		Thus, in the weak-pulling regime, the frequency shift associated with the weakly excited coordinate depends linearly on the Duffing shift of the driven coordinate. 

		To probe this coupling experimentally, we coherently drive the predominantly $x$-like mode and sweep $\omega_\mathrm{d}$ downward across its resonance, as described in Sec.~\ref{sec:nonlinear_response}. During the sweep, the driven mode follows the high-amplitude Duffing branch and acquires an increasing amplitude-dependent frequency shift. Although the $y$-like mode is far detuned from the drive, its frequency shifts downward as $A_x$ increases, as shown in Fig.~\ref{fig:ringdown}(c). For $\vert\Delta_+\vert/2\pi<\SI{600}{\hertz}$, the measured shifts of the eigenfrequencies, $\delta \omega_-$ and $\delta \omega_+$, satisfy the linear relation in Eq.~\eqref{eq:pulling_linear_relation}, as shown in Fig.~\ref{fig:ringdown}(d). From the fitted slope $\delta \omega_-/\delta \omega_+=0.225(2)$, we extract the effective nonlinear coupling coefficient $g_2/(2\pi)^2=-0.5(1)\times 10^{22}~\mathrm{Hz^2/m^2}$, which incorporates the mode hybridization. For $\vert\Delta_+\vert/2\pi>\SI{600}{\hertz}$, deviations from the linear relation emerge. These deviations may indicate the increasing importance of higher-order nonlinearities or the breakdown of the weak-pulling and single-mode approximations at large oscillation amplitudes.

\subsection{Comb-formation model}\label{app:freq_comb}
	The primary frequency comb emerges from nondegenerate parametric excitation of the two hybridized translational modes, followed by nonlinear frequency mixing. We first derive the resonant dynamics responsible for parametric excitation and sum-phase locking of the carrier tones and subsequently discuss the formation of the accompanying sidebands. We then extend the model to include the $\gamma$ libration, whose nonlinear excitation provides a mechanism for the finer spectral structure of the full comb and the stabilization of its spacing. 

	\subsubsection{Hybrid-mode dynamics}
		We consider the regime in which only the $x$ coordinate is directly driven. For the analytical treatment, we retain only the self-Duffing nonlinearities of $x$ and $y$ and neglect the additional nonlinear intermodal interactions characterized in Sec.~\ref{app:nonlinear_coupling}, since they produce only quantitative corrections to the nonlinear frequency shifts and are not required to capture the leading excitation and phase-locking mechanism. 

		We focus on the sum-frequency resonance $\omega_\mathrm{d}\approx \omega_+ + \omega_-$, which enables nondegenerate parametric three-wave mixing between the hybrid modes. To make this interaction explicit, we introduce the normal coordinates $q_\pm$,
		\begin{equation}\label{eq:app_eigenmodes_rotation}
			\begin{pmatrix}x \\ y\end{pmatrix}=\begin{pmatrix}\cos\theta & -\sin\theta \\ \sin\theta & \cos\theta \end{pmatrix}
			\begin{pmatrix}
			q_+ \\ q_-\end{pmatrix},
		\end{equation}
		where the mixing angle $\theta$ is defined as in Eq.~\eqref{eq:mixed_angle}. This transformation diagonalizes the conservative linear dynamics, yielding the hybrid eigenfrequencies $\omega_\pm$. Near the sum-frequency resonance, the relevant projected parametric term couples $q_+$ and $q_-$. 
		 Retaining this term, the projected coherent drive, and the dominant self-Duffing nonlinearities gives 
		\begin{equation}
			\ddot q_\pm=-\Gamma_\pm\dot q_\pm-\omega_\pm^2 q_\pm+\mu\cos(\omega_\mathrm{d}t)\,q_\mp-\lambda_\pm q_\pm^3+\eta_\pm\cos(\omega_\mathrm{d}t).
			\label{eq:eom_q_reduced}
		\end{equation}
		Here, the effective parameters read
		\begin{subequations}
			\begin{align}
				\mu = -\varepsilon \cos\theta \sin\theta,\quad \eta_+ = \eta \cos\theta,\quad \eta_-=-\eta\sin\theta,\quad\lambda_+ = \lambda_x \cos^4\theta+\lambda_y \sin^4\theta,\quad \lambda_-=\lambda_x \sin^4\theta+\lambda_y \cos^4\theta,
			\end{align}
			and the effective damping rates are
			\begin{align}
				\Gamma_+ = \Gamma_x\cos^2\theta+\Gamma_y\sin^2\theta,\qquad
				\Gamma_- = \Gamma_x\sin^2\theta+\Gamma_y\cos^2\theta.
			\end{align}
		\end{subequations}
		The sign of $\mu$ follows from the convention chosen in Eq.~\eqref{eq:app_eigenmodes_rotation} and can equivalently be absorbed into the phase of the parametric drive. For simplicity, off-diagonal damping, the mixed nonlinear terms generated by projecting the coordinate Duffing nonlinearities, and the off-resonant diagonal parametric modulation terms have been omitted from the analytical reduction. These contributions do not alter the linear parametric-instability threshold or the leading sum-phase-locking mechanism, although they provide quantitative corrections to the finite-amplitude carrier-frequency shifts. 

		To isolate the resonant slow-envelope dynamics, we write
		\begin{equation}
			q_\pm(t)=\dfrac{1}{\sqrt{2}}\left[\beta_\pm(t)\mathrm{e}^{-i\omega_\pm t}+\beta_\pm^*(t)\mathrm{e}^{i\omega_\pm t}\right],
			\qquad
			|\dot{\beta}_\pm| \ll \omega_\pm |\beta_\pm|.
		\end{equation}
		Retaining only terms that vary slowly near the sum-frequency resonance gives
		\begin{align}
			\dot{\beta}_\pm=&-\dfrac{\Gamma_\pm}{2}\beta_\pm+i\mu_\pm \beta_\mp^* \mathrm{e}^{-i\Delta t}-iD_\pm |\beta_\pm|^2\beta_\pm,
			\label{eq:app_eom_beta}
		\end{align}
		where
		\begin{equation}
			\mu_\pm = \dfrac{\mu}{4\omega_\pm}, \qquad
			D_\pm = \dfrac{3\lambda_\pm}{4\omega_\pm}, \qquad
			\Delta = \omega_\mathrm{d}-(\omega_+ + \omega_-).
		\end{equation}
		Equation~\eqref{eq:app_eom_beta} directly describes nondegenerate parametric three-wave mixing: the term proportional to $\beta_\mp^\ast$ converts the parametric drive into a correlated pair of hybrid-mode oscillations and becomes resonant for $\Delta\approx 0$, where the coupling term is nearly time-independent. The Duffing terms produce the amplitude-dependent carrier-frequency shifts that enter the phase-locking dynamics derived below. Their additional role in generating the comb sidebands beyond the leading resonant approximation is discussed qualitatively after the locking analysis. The projected coherent drive would contribute terms proportional to $\mathrm{exp}[-i(\omega_\mathrm{d}-\omega_\pm)t]$. Near $\omega_\mathrm{d}- \omega_\pm\approx \omega_\mp$, these terms vary rapidly in the rotating frames and are thus omitted from the resonant slow-envelope equations. In the full dynamics, however, the coherent drive produces the prominent spectral tone at $\omega_\mathrm{d}$. This tone provides an additional seed for nonlinear mixing with the parametrically generated carrier tones and thereby contributes to the formation of the comb sidebands.

	\subsubsection{Parametric-instability threshold and phase locking}
		To derive the condition for parametric excitation, we write the complex amplitudes as 
		\begin{equation}
			\beta_\pm(t)=|\beta_\pm(t)| \mathrm{e}^{ i\varphi_\pm(t)},
			\label{eq:app_beta_phase}
		\end{equation}
		and define the phase mismatch between the parametric drive and the carrier pair as 
		\begin{equation}
			\Phi(t) = \varphi_+(t)+\varphi_-(t)+\Delta\,t.
			\label{eq:app_phase_sum}
		\end{equation}
		Substituting this representation into Eq.~\eqref{eq:app_eom_beta} and separating the amplitude and the phase dynamics gives 
		\begin{subequations}
			\begin{equation}
				\dfrac{\mathrm{d}}{\mathrm{d}t}|\beta_\pm| = -\dfrac{\Gamma_\pm}{2} |\beta_\pm|+\mu_\pm |\beta_\mp|\sin\Phi,
			\end{equation}
			and
			\begin{equation}
				\dot{\varphi}_\pm = \mu_\pm \dfrac{|\beta_\mp|}{|\beta_\pm|}\cos\Phi-D_\pm|\beta_\pm|^2.
				\label{eq:app_dot_phi_pm}
			\end{equation}
		\end{subequations}
		The amplitude equations describe the competition between parametric gain and dissipation, while the phase equations contain the parametric interaction and the Duffing-induced frequency shifts. 
		Summing the phase equations gives an Adler-type phase equation 
		\begin{subequations}
			\begin{equation}
				\dot{\Phi}=\Delta_\mathrm{eff}+A_\mathrm{eff}\cos\Phi,
				\label{eq:eom_phi}
			\end{equation}
			where
			\begin{equation}
				\Delta_\mathrm{eff}=\Delta-(D_+|\beta_+|^2+D_-|\beta_-|^2), 
			\end{equation}
			and
			\begin{equation}
				A_\mathrm{eff}=\mu_+\dfrac{|\beta_-|}{|\beta_+|}+\mu_-\dfrac{|\beta_+|}{|\beta_-|}.
			\end{equation}
		\end{subequations}
		Here, $\Delta_\mathrm{eff}$ contains both the drive detuning and the Duffing-induced carrier-frequency shifts, while $A_\mathrm{eff}$ quantifies the effective parametric coupling. For fixed amplitudes, the phase equation admits a stable fixed point when 
		\begin{equation}
			|\Delta_\mathrm{eff}|<|A_\mathrm{eff}|.
		\end{equation}
		A stationary phase-locked solution satisfies both $\mathrm{d}|\beta_\pm|/\mathrm{d}t=0$ and $\dot{\Phi}=0$. The amplitude equations then give
		\begin{align}
			\dfrac{|\beta_+|}{|\beta_-|} &= \sqrt{\dfrac{\Gamma_- \mu_+}{\Gamma_+ \mu_-}},\qquad
			\sin\Phi= \sqrt{\dfrac{\Gamma_+\Gamma_-}{4\mu_+\mu_-}}.
			\label{eq:app_ampRatio_phase}
		\end{align}
		A nonzero stationary solution therefore requires
		\begin{equation}
			4|\mu_+\mu_-|>\Gamma_+\Gamma_-.
			\label{eq:app_instability_threshold}
		\end{equation}
		At exact sum-frequency resonance, $\Delta=0$, this condition is the nondegenerate parametric-instability threshold of the zero-amplitude state. Above threshold, the parametric gain overcomes dissipation, and the carrier amplitudes grow until the Duffing-induced frequency shifts detune the system sufficiently to establish a stationary-amplitude state. At finite detuning, $\Delta\neq0$, it remains a necessary condition for a stationary parametrically excited state, while the existence and stability of the locked solution additionally depend on the phase dynamics.

		Using the phase-locking condition $\dot{\Phi}=0$ in Eq.~\eqref{eq:app_dot_phi_pm} together with the stationary amplitude ratio~\eqref{eq:app_ampRatio_phase}, the stationary amplitude is
		\begin{equation}
			|\beta_+|^2=\dfrac{\Gamma_- \mu_+\left[\Delta+\left(\sqrt{\Gamma_+\mu_+ \mu_-/\Gamma_-} + \sqrt{\Gamma_-\mu_- \mu_+/\Gamma_+ }\right)\cos\Phi\right]}{\Gamma_- \mu_+D_+ + \Gamma_+ \mu_-D_-},
			\label{eq:limitCycle_amp_ratio}
		\end{equation}
		which reflects the balance among drive detuning, nonlinear frequency pulling, parametric coupling, and dissipation. Defining the resulting carrier-frequency shift from the linear eigenfrequencies as $\delta_\pm=\Omega_\pm-\omega_\pm=-\dot{\varphi}_\pm$, the phase equations give
		\begin{equation}
			\delta_\pm=D_\pm |\beta_\pm|^2-\mu_\pm\dfrac{|\beta_\mp|}{|\beta_\pm|}\cos\Phi.
		\end{equation}
		At the phase-locked fixed point, $\dot{\Phi}=0$ implies $\delta_++\delta_-=\Delta$ and $\Omega_++\Omega_-=\omega_\mathrm{d}$. Thus, the individual carrier-frequency shifts contain contributions from both Duffing frequency pulling and the parametric interaction, while their sum equals the detuning, ensuring that the sum of the carrier frequencies remains locked to the drive. At $\Delta=0$, the two shifts are equal and opposite. The resulting parametrically generated, phase-correlated carrier pair provides the background from which nonlinear mixing produces the primary frequency comb.

	\subsubsection{Nonlinear frequency mixing and sideband formation}
		The resonant slow-envelope dynamics derived above describe the onset of parametric excitation and the sum-phase locking of the carrier pair at frequencies $\Omega_\pm$, but omit the rapidly oscillating contributions discarded within the RWA. Restoring the off-resonant coherent-drive contribution additionally produces the pronounced spectral tone at $\omega_\mathrm{d}$. Although this coherent response does not determine the parametric-instability threshold, it provides an additional tone that can participate in nonlinear mixing. 

		The Duffing restoring-force terms $\lambda_x x^3$ and $\lambda_y y^3$ in Eq.~\eqref{eq:eom_xy} generate both self- and cross-mode cubic contributions when expressed in the hybrid basis and thereby mediate four-wave mixing among the carrier tones and the coherent response at $\omega_\mathrm{d}$. The lowest-order mixing products of the carrier pair occur at $2\Omega_+-\Omega_-$ and $2\Omega_--\Omega_+$, corresponding, respectively, to sidebands one frequency difference $\omega_\mathrm{comb}=|\Omega_+-\Omega_-|$ above the upper carrier and below the lower carrier. Mixing involving the coherent drive similarly generates components around $\omega_\mathrm{d}$. In particular, the frequencies $2\Omega_+$ and $2\Omega_-$ lie symmetrically about $\omega_\mathrm{d}$, each separated from it by $\omega_\mathrm{comb}$ in the sum-locked regime. These newly generated components subsequently participate in further four-wave-mixing processes, producing a cascade of combination tones at frequencies $j\Omega_++k\Omega_-$ with $j,k\in\mathbb{Z}$. 

		Assuming $\Omega_+>\Omega_-$, the combination frequencies can be written as 
		\begin{equation}
		    \Omega_{jk}=\dfrac{j+k}{2}\omega_\mathrm{d}+\dfrac{j-k}{2}\omega_\mathrm{comb}.
		\end{equation}
		Thus, the spectral components can be grouped into families with fixed $j+k$, \textit{e.g.}, the family with $j+k=1$ contains the carrier tones $\Omega_\pm$ and their associated sidebands, whereas the family with $j+k=2$ contains the coherent-drive tone $\omega_\mathrm{d}=\Omega_++\Omega_-$ and its sidebands. Neighboring components within each family are separated by the primary-comb spacing $\omega_\mathrm{comb}$. Cascaded Duffing-induced four-wave-mixing therefore produces the equidistant sideband ladders observed around the carrier tones and the coherent drive, forming the primary frequency comb. Its spacing is set by the drive-dependent nonlinear carrier frequencies $\Omega_\pm$, while the amplitudes of the individual comb teeth depend on the detunings, damping rates, and nonlinear coupling strengths and are not evaluated within the reduced resonant model.

	\subsubsection{Resonance condition for sideband merging}\label{app:sideband_merge}
		Spectral components belonging to different combination-tone families can coincide when the comb spacing becomes commensurate with the drive frequency. At such degeneracies, multiple nonlinear mixing pathways contribute to the same spectral component and may enhance or redistribute its spectral weight. As a relevant example, we consider an upper sideband associated with the lower-frequency carrier at $\Omega_-+n\omega_\mathrm{comb}$, with $n\geq1$, and a lower sideband of the coherent-drive tone at $\omega_\mathrm{d}-m\omega_\mathrm{comb}$, with $m\geq0$. These components coincide when $ \Omega_- + n \omega_\mathrm{comb}= \omega_\mathrm{d} - m \omega_\mathrm{comb}$.
		Using $\Omega_- = (\omega_\mathrm{d} - \omega_\mathrm{comb})/2$, the degeneracy condition becomes 
		\begin{equation}
			\left[2(n + m) - 1\right]\omega_\mathrm{comb}= \omega_\mathrm{d}.
		\end{equation}
		 Defining $p=n+m-1$ therefore gives
		\begin{equation}
			\omega_\mathrm{comb} = \dfrac{\omega_\mathrm{d}}{2p+1},\qquad p\in\mathbb{N},
			\label{eq:locking_condition}
		\end{equation}
		showing that sidebands from these two families merge when the ratio $\omega_\mathrm{d}/\omega_\mathrm{comb}$ is an odd integer. In particular, the experimentally observed ratio $\omega_\mathrm{d}/\omega_\mathrm{comb}=15$ corresponds to $p=7$. More generally, crossings between other combination-tone families can occur at rational values of $\omega_\mathrm{d}/\omega_\mathrm{comb}$.

		The commensurability condition alone, however, does not imply dynamical locking of $\omega_\mathrm{comb}$ and the observed extended plateau, motivating the inclusion of the $\gamma$ libration. 
	\subsubsection{Coupling to $\gamma$ librations}\label{app:coupling_gamma}
		For completeness, we specify the three-mode model used in the numerical simulations. We integrate Eqs.~\eqref{eq:eom_xy} and \eqref{eq:eom_gamma}, supplementing the $x$ equation with the reciprocal linear $x$-$\gamma$ coupling and the reciprocal nonlinear terms implied by the interaction potential in Eq.~\eqref{eq:V_nl_reduced}. With the coefficient normalization defined in Sec.~\ref{sec:nonlinear_expansion}, this gives an additional contribution of 
		\begin{equation}
		    a_\gamma = -\dfrac{I_\gamma}{M}\left(g_\gamma \gamma + \chi\gamma^2+2\zeta x\gamma\right)
		\end{equation}
		on the right-hand side of the equation of motion of $x$. Direct coupling between $y$ and $\gamma$, as well as higher-order interactions, are omitted from the minimal model. The threshold for the $\chi$-induced parametric excitation of $\gamma$ is determined by the amplitude of the relevant primary-comb component, the damping rate $\Gamma_\gamma$, and the detuning from $2\omega_\gamma$. 

		\textit{Numerical implementation.---}
		We numerically integrate the deterministic three-mode equations in dimensionless units. Time is rescaled as $\tilde{t}=\omega_\mathrm{ref}t$, with the reference frequency chosen as $\omega_\mathrm{ref}/2\pi=16\,\mathrm{kHz}$. The generalized coordinates are additionally rescaled such that the mass and amplitude normalization factors are absorbed into the effective dimensionless coupling coefficients. The dimensionless parameters used in the simulation are summarized in Table~\ref{tab:simulation_parameters}. They are chosen phenomenologically to reproduce the hierarchy of nonlinear dynamical regimes and are not obtained from a quantitative fit to the experimental parameters. Accordingly, the effective coordinates of the reduced numerical model are not identified one-to-one with the spatial mode polarizations determined experimentally. For the parameter set used here, the driven coordinate projects predominantly onto the lower-frequency hybrid branch. The comparison is therefore intended to capture the nonlinear spectral organization rather than the measured branch-specific mode polarizations.
		\begin{table}[b]
			\caption{Dimensionless parameters used in the three-mode numerical simulation. Here, $\tilde{k}_{xy}=2g\omega_0/\omega_\mathrm{ref}^2$ is the dimensionless linear coupling coefficient.}
			\label{tab:simulation_parameters}
			\centering
			\begin{tabular}{|c|c|}
				\hline
				Parameter & Value\\
				\hline
				$(\tilde{\omega}_x,\tilde{\omega}_y, \tilde{\omega}_\gamma)$ & $(0.883,1,0.304)$\\
				$(\tilde{\Gamma}_x,\tilde{\Gamma}_y, \tilde{\Gamma}_\gamma)$ & $(10^{-4},10^{-4},10^{-4})$\\
				$(\tilde{\lambda}_x,\tilde{\lambda}_y, \tilde{\lambda}_\gamma)$ & $(-0.8,-0.1,0.1)$\\
				$\tilde{k}_{xy}$ & $0.03$\\
				$\tilde{g}_\gamma$ & $0.05$\\
				$\tilde{\zeta}$ & $0.09$\\
				$\tilde{\chi}$ & $0.195$\\
				$\tilde{\varepsilon}$ & $0.156$\\
				$\tilde{\eta}$ & $0.25$\\
				\hline
			\end{tabular}
		\end{table}

\begin{figure}[t!]
		    \centering
		    \includegraphics[width=\columnwidth]{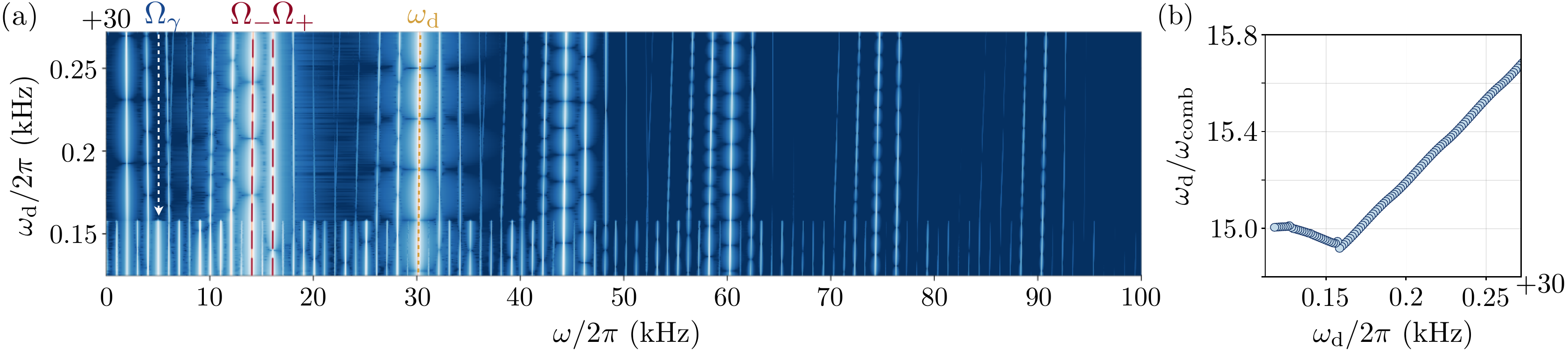}
		    \caption{Numerical simulation of the downward drive-frequency sweep. 
		    (a) Spectrum of the $x$ coordinate acquired during the downward sweep, obtained from integrating the extended nonlinear version of Eqs.~\eqref{eq:eom_xy} and \eqref{eq:eom_gamma} for parameters specified in Table~\ref{tab:simulation_parameters}. 
		    (b) Ratio $\omega_\mathrm{d}/\omega_\mathrm{comb}$ as a function of $\omega_\mathrm{d}$, where $\omega_\mathrm{comb}$ is extracted from the two predominantly translational carrier frequencies highlighted by the red dashed lines. }
		    \label{fig:gamma_sweep}
		\end{figure}

		The two predominantly translational eigenfrequencies $\tilde{\omega}_\pm$ are obtained by diagonalizing the linearized stiffness matrix. The dimensionless instantaneous drive frequency $\tilde{\omega}_\mathrm{d}(\tilde{t})$ is swept linearly from $\tilde{\omega}_\mathrm{d}^{\,\mathrm{start}}=\tilde{\omega}_++\tilde{\omega}_-+0.0075$ to $\tilde{\omega}_\mathrm{d}^{\,\mathrm{end}}=\tilde{\omega}_++\tilde{\omega}_- -0.0025$ over a dimensionless time interval $\tilde{T}_{\mathrm{sweep}}=5\times10^6$, with the drive phase defined by $\mathrm{d}\phi_\mathrm{d}/\mathrm{d}\tilde{t}=\tilde{\omega}_\mathrm{d}(\tilde{t})$. The dynamics are initialized at rest and evolved without stochastic forcing. The simulated spectrogram is obtained from the $x$-coordinate trajectory using consecutive, overlapping time windows. The unprocessed relative spectrum is shown in Fig.~\ref{fig:gamma_sweep}(a). To improve the visibility of weak, narrow comb lines over the large dynamic range of the simulation, we use a peak-enhanced representation in Fig.~\ref{fig3:gamma}(a). At each drive frequency, the spectral weight is compared with the local background at nearby frequencies. Components that do not form a sufficiently prominent local peak are suppressed, while the color of the retained components is determined by their original relative spectral power. A short averaging over neighboring sweep points suppresses isolated numerical features.

		\textit{Simulation results.---}
		Figure~\ref{fig:gamma_sweep}(a) shows the unprocessed simulated spectrum of the $x$-coordinate during the downward sweep of $\omega_{\mathrm d}$. Before excitation of the $\gamma$ libration, the spectrum is dominated by the two parametrically generated carrier tones and the accompanying primary-comb sidebands. As the drive frequency decreases, a nonlinear mixing component at $ 5\omega_\mathrm{comb}$ approaches twice the effective $\gamma$-mode frequency, satisfying the parametric-resonance condition. Once the corresponding modulation amplitude exceeds the parametric-instability threshold, the librational amplitude increases sharply. We denote the drive-dependent frequency of the resulting librational carrier by $\Omega_\gamma$. The linear and nonlinear couplings to the $x$ coordinate then transfer the resulting $\gamma$-mode components into the $x$ spectrum, producing an additional family of sidebands at approximately
		\begin{equation}
			\Omega_\gamma+m\omega_\mathrm{comb},\qquad m\in\mathbb Z.
			\label{eq:gamma_sideband_family}
		\end{equation}
		In the phase-locked regime, where $\Omega_\gamma=5\omega_\mathrm{comb}/2$, this family lies approximately halfway between neighboring primary-comb teeth. Its emergence therefore produces the denser spectral structure associated with the full-comb regime.

		To characterize the accompanying evolution of the primary comb, we extract the two carrier frequencies $\Omega_\pm$ throughout the sweep and calculate $\omega_\mathrm{comb}=|\Omega_+-\Omega_-|$, as shown in Fig.~\ref{fig:gamma_sweep}(b). Before the onset of the $\gamma$-mode instability, the ratio $\omega_{\mathrm d}/\omega_\mathrm{comb}$ varies continuously with the drive frequency. Once the $\gamma$ libration acquires a finite amplitude, it renormalizes the translational carrier frequencies and produces a plateau near $\omega_{\mathrm d}/\omega_\mathrm{comb}\approx 15$. The simulation therefore qualitatively reproduces both the interleaved spectral structure and the stabilization of the primary-comb spacing observed experimentally. Because the coupling coefficients are chosen phenomenologically, the comparison is intended to capture the physical mechanism and spectral organization rather than the absolute instability threshold or the amplitudes of individual comb teeth.

\twocolumngrid

\bibliography{Main.bib}

\end{document}